\documentclass[a4paper,onecolumn,11pt,unpublished]{quantumarticle}
\pdfoutput=1
\usepackage[utf8]{inputenc}
\usepackage[english]{babel}
\usepackage[T1]{fontenc}
\usepackage{amsmath}
\usepackage{hyperref}
\usepackage[numbers]{natbib}
\usepackage{amsfonts,graphicx,subfigure,color}

\usepackage{tikz}
\usepackage{lipsum}
\usepackage{notes2bib}

\begin{document}

\title{Feedback-based quantum algorithms for ground state preparation}

\author{James~B. Larsen}
\affiliation{Department of Mathematics, University of Michigan, Ann Arbor, MI 48109, USA.}
\affiliation{Quantum Algorithms and Applications Collaboratory, Sandia National Laboratories, Albuquerque, New Mexico 87185, USA.}
\affiliation{Department of Mathematics, Brigham Young University, Provo, UT 84604, USA.}
\orcid{0000-0002-0777-440X}
\email{jblarse@sandia.gov}
\author{Matthew~D. Grace}
\affiliation{Quantum Algorithms and Applications Collaboratory, Sandia National Laboratories, Livermore, California 94550, USA.}
\orcid{0000-0001-5842-7347}
\author{Andrew~D. Baczewski}
\affiliation{Quantum Algorithms and Applications Collaboratory, Sandia National Laboratories, Albuquerque, New Mexico 87185, USA.}
\author{Alicia~B. Magann}
\affiliation{Quantum Algorithms and Applications Collaboratory, Sandia National Laboratories, Albuquerque, New Mexico 87185, USA.}
\orcid{0000-0002-1402-3487}
\email{abmagan@sandia.gov}
\maketitle

\begin{abstract}
The ground state properties of quantum many-body systems are a subject of interest across chemistry, materials science, and physics. Thus, algorithms for finding ground states can have broad impacts. Variational quantum algorithms are one class of ground state algorithms that has received significant attention in recent years. These algorithms utilize a hybrid quantum-classical computing framework to prepare ground states on quantum computers. However, this requires solving a classical optimization problem that can become prohibitively expensive in high dimensions. Here, we develop formulations of feedback-based quantum algorithms for ground state preparation that can be used to address this challenge for two broad classes of Hamiltonians: Fermi-Hubbard Hamiltonians, and molecular Hamiltonians represented in second quantization. Feedback-based quantum algorithms are optimization-free; in place of classical optimization, quantum circuit parameters are set according to a deterministic feedback law derived from quantum Lyapunov control principles. This feedback law guarantees a monotonic improvement in solution quality with respect to the depth of the quantum circuit. A variety of numerical illustrations are provided that analyze the convergence and robustness of feedback-based quantum algorithms for these problem classes.
\end{abstract}

\section{Introduction} \label{JBL:sec:intro}

The ground state properties of quantum many-body systems are related to many features of interest that are studied using quantum simulation algorithms. Thus, algorithms for preparing ground states are one of the central topics of interest in quantum simulation research \cite{poulin2009preparing, tubman2018postponing, ge2019faster, lin2020near, wang2023ground}. While we expect the problem of finding generic ground states to be hard, even for quantum computers \cite{Kitaev2002May, kempe2006complexity}, physical intuition suggests that systems that efficiently equilibrate in nature may also be amenable to efficient ground state preparation on a quantum computer. Recent progress in the development of quantum computers has given rise to the noisy intermediate-scale quantum (NISQ) era \cite{JBL:nisq}. This progress has motivated substantial research exploring applications of NISQ devices in a variety of areas, including quantum simulation \cite{cerezo2021variational, RevModPhys.94.015004}. In particular, the prospect of utilizing NISQ devices to find ground states of chemical and materials systems is currently receiving significant attention. To this end, the Variational Quantum Eigensolver (VQE) \cite{JBL:peruzzo2014variational} has been developed as a hybrid quantum-classical algorithm for this task that may hold promise in the near-term. However, a practical challenge associated with VQE is the need to classically optimize over a set of quantum circuit parameters in order to minimize an objective function, a task whose complexity can increase rapidly with the dimension of the search space. 

\begin{figure}[t]
\begin{center}
\scalebox{0.25}{\includegraphics{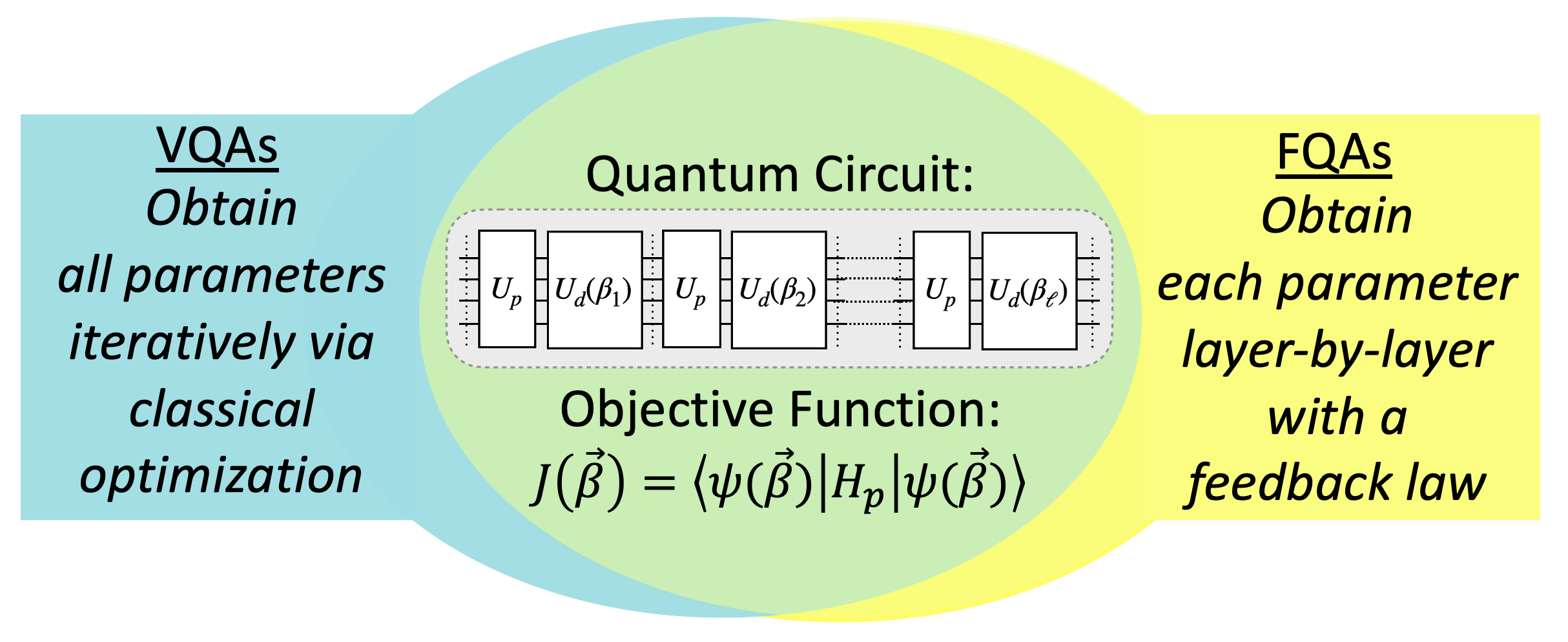}} \caption{Venn diagram illustrating conceptual similarities and differences between variational quantum algorithms (VQAs) and feedback-based quantum algorithms (FQAs). As depicted in the center, VQAs and FQAs can be formulated using the same objective function $J$ and quantum circuit structure. However, they approach the task of minimizing $J$ in fundamentally different ways. In particular, VQAs operate by classically optimizing over all parameters $\vec{\beta}$ (left). FQAs replace the classical optimization routine with a feedback law that is used to set the quantum circuit parameters one at a time, layer by layer (right).
}
\label{figure1}
\end{center}\end{figure}

We investigate an alternative formulation for finding ground states that is optimization-free, thereby avoiding this complexity. Instead of classically optimizing over a parameterized circuit, this formulation utilizes a feedback law to sequentially set the quantum circuit parameter values layer-by-layer, conditioned on feedback from qubit measurements from the prior layer. The feedback law is based on the theory of quantum Lyapunov control \cite{JBL:lyapunovsurvey, JBL:1272601}, and is designed to ensure that the value of the objective function monotonically decreases with respect to circuit depth. This represents a new application for Feedback-based Quantum Algorithms (FQAs), generalizing the Feedback-based ALgorithm for Quantum OptimizatioN (FALQON) \cite{JBL:falqon, JBL:longfalqon} that was developed for applications in combinatorial optimization. Figure~\ref{figure1} highlights the key difference between FQAs and Variational Quantum Algorithms (VQAs) like VQE. We focus here on the development of FQAs for finding ground states of systems of interacting electrons (fermions), specifically centering our attention on the Fermi-Hubbard model and on molecular Hamiltonians formulated in second quantization.

The remaining sections are organized as follows. In Sec.~\ref{JBL:Sec:prelims}, we review relevant theory, including a description of electronic structure Hamiltonians, the Fermi-Hubbard model, VQE, quantum Lyapunov control, and FALQON. Next, in Sec.~\ref{JBL:sec:apply}, we combine these preliminaries to formulate FQAs for ground state preparation. We then present a variety of numerical results and analyses considering different instances of the Fermi-Hubbard model and molecular Hamiltonians in Secs.~\ref{JBL:Sec:results} and \ref{sec:molham} respectively. We then conclude with a discussion in Sec.~\ref{Sec:Discussion} and possible future avenues of research in Sec.~\ref{JBL:Sec:conclu}.

\section{Preliminaries}
\label{JBL:Sec:prelims}
In this section, we review technical details relevant to the motivation and development of FQAs and their application to ground state preparation.

\subsection{The Electronic Structure Hamiltonian}
\label{sec:molhamint}
The generic second-quantized electronic structure Hamiltonian takes the form
\begin{equation}
H_{MH} = H_1 + H_2,~\label{eq:electronic_structure_hamiltonian}
\end{equation}
with
\begin{equation}
H_1 = \sum_{pq} h_{pq} \hat{a}_p^\dagger \hat{a}_q
\end{equation}
and
\begin{equation}
H_2 = \frac{1}{2} \sum_{pqrs} h_{pqrs} \hat{a}_p^\dagger \hat{a}_q^\dagger \hat{a}_r \hat{a}_s.
\end{equation}
The operator $\hat{a}_{p}$ ($\hat{a}_{p}^\dagger$) annihilates (creates) a fermion in the $p$th spin orbital, which is defined with respect to the molecular orbitals that diagonalize a mean-field Hamiltonian, the sites on a lattice model, or some other convenient association relevant to the physical system being modeled. Meanwhile, the coefficients $h_{pq}$ and $h_{pqrs}$ are referred to as the one- and two-electron integrals, respectively~\cite{szabo2012modern}.
With the assistance of quantum chemistry software packages, these integrals are typically evaluated in a basis of atomic orbitals that are subsequently transformed into a basis of molecular orbitals in which Eq.~\eqref{eq:electronic_structure_hamiltonian} is expressed.
Hamiltonians of this form can be used to describe a vast array of chemical systems, from small molecules to solids and beyond. Consequently, there is significant interest in algorithms for preparing their ground states on quantum computers~\cite{cao2019quantum,mcardle2020quantum} using techniques like VQE (see Sec. \ref{sec:vqe}). In this work, we develop FQAs to eliminate the expensive optimization loop in VQE for electronic structure Hamiltonians by first focusing on a specific class of instances: the Fermi-Hubbard model, and then broadening to more generic formulations relevant to small molecules.

\subsection{The Fermi-Hubbard Model}
The Fermi-Hubbard model captures some of the basic phenomenology of strongly correlated systems without many of the details that complicate first-principles models of strongly correlated molecules or solids. It has been the subject of intense interest since its introduction in 1963 \cite{JBL:hubbard}. In particular, the Fermi-Hubbard model describes the behavior of fermions hopping between lattice sites and experiencing repulsive on-site interactions, as illustrated pictorially in Fig.~\ref{fig:fh}(a). Despite its simplicity, the Fermi-Hubbard model can represent a variety of orderings and certain details of its phase diagram remain an open topic of research \cite{JBL:arovas22hubbard, JBL:qin22hubbard}. As an approximation to more detailed models of numerous material systems, it is difficult to unambiguously constrain the model's phase diagram through experiments, even with bespoke quantum simulators \cite{cocchi2016equation, mazurenko2017cold, brown2019bad}. Thus, simulations provide a natural path for studying the model. At zero temperature, this involves approximating the ground state of the model for finite-size instances and extrapolating to the thermodynamic limit.
The difficulty of precisely constraining solutions to various instances of the model highlights a fundamental challenge in computational physics: classical computers force us to choose between exact algorithms, requiring resources that scale exponentially with the number of lattice sites, and approximate algorithms that are efficient but without easily quantifiable errors. The exponential cost of exact classical algorithms is nicely highlighted by one of the largest exact simulations done on a particular variant of the 2D Fermi-Hubbard model, with 22 sites and 17 electrons \cite{JBL:yamada200516}. This challenge motivates the use of quantum computers. To approach the thermodynamic limit while preserving quantifiable error, it may be that quantum computers can provide advantages.

The Fermi-Hubbard Hamiltonian contains two terms, 
\begin{equation}
    \label{JBL:FHHam}
    H_{FH} = T + V,
\end{equation}
where $T$ denotes the kinetic term associated with fermions hopping between lattice sites and forms the noninteracting portion of the Hamiltonian:
\begin{equation}
    \label{JBL:kinetic}
    T = - \sum_{\langle m,n\rangle} t_{m,n} (\hat{a}_{m\uparrow}^\dagger\hat{a}_{n\uparrow} + \hat{a}_{n\uparrow}^\dagger\hat{a}_{m\uparrow} + \hat{a}_{m\downarrow}^\dagger\hat{a}_{n\downarrow} + \hat{a}_{n\downarrow}^\dagger\hat{a}_{m\downarrow}),
\end{equation}
where $\langle m,n \rangle$ labels adjacent sites in the lattice, $t_{m,n}$ denotes the tunneling amplitude associated with hopping from site $m$ to site $n$, and $\hat{a}_{m\uparrow}^\dagger$ ($\hat{a}_{m\uparrow}$) denotes the fermionic creation (annihilation) operator associated with the $\uparrow$ spin orbital of lattice site $m$. For future analysis, we set the tunneling amplitudes $t_{m,n}$ from Eq.~(\ref{JBL:kinetic}) to be equal to the same value, $\tau$, for all ${\langle m, n\rangle}$-pairs. Meanwhile, $V$ in Eq.~\eqref{JBL:FHHam} represents the local potential energy at each site that is due to the on-site interaction between fermions:
\begin{equation}
    \label{JBL:potential}
    V = U \sum_{k=1}^N \hat{n}_{k\uparrow} \hat{n}_{k\downarrow},
\end{equation}
with $U$ representing the on-site repulsive interaction strength, where $\hat{n}_{k\uparrow} = \hat{a}_{k\uparrow}^\dagger \hat{a}_{k\uparrow}$.  In general, there are very few scenarios in which the Fermi-Hubbard model is exactly solvable \cite{JBL:lieb2003one} or amenable to systematically improvable numerical methods \cite{hirsch1985two,wu2005sufficient}.

\subsection{The Variational Quantum Eigensolver} \label{sec:vqe}

VQAs seek to minimize a cost function $J$ through a combination of quantum and classical computation \cite{cerezo2021variational, RevModPhys.94.015004}. On the quantum computer, a parameterized quantum circuit is utilized to evaluate $J(\vec{\beta})$ as a function of quantum circuit parameters $\vec{\beta}$. Meanwhile, a classical computer is used to search over the associated parameter space in order to identify the parameter configuration $\vec{\beta}_{min}$ corresponding to a minimum of $J$. VQE leverages this hybrid quantum-classical framework to seek the ground state of a many-body problem Hamiltonian, $H_p$, by setting $J(\vec{\beta})=\langle\psi(\vec{\beta})|H_p|\psi(\vec{\beta})\rangle$, such that finding the global minimum corresponds to preparing the ground state of $H_p$ \cite{JBL:peruzzo2014variational, McClean_2016}. VQE proceeds by first initializing a set of qubits in a state $|{\psi_0}\rangle$ and then applying a parameterized unitary (associated with a desired ansatz) to $|{\psi_0}\rangle$, yielding $U(\vec{\beta})|{\psi_0}\rangle = |{\psi(\vec{\beta})}\rangle$ for some initialization of the parameter vector $\vec{\beta}$. Next, the expectation value of $H_p$ under the state $|{\psi(\vec{\beta})}\rangle$ is estimated via qubit measurements to obtain $J(\vec{\beta})$. The classical optimizer then updates $\vec{\beta}$ in an effort to minimize $J$. This process is repeated to suitable convergence of $J(\vec{\beta})$ \cite{JBL:VQEconvergence}.

There has been significant interest in using VQE to prepare ground states of the electronic structure Hamiltonian and the Fermi-Hubbard Hamiltonian on quantum computers. Regarding the former, the VQE has been demonstrated in quantum hardware for a range of different molecular Hamiltonians and quantum circuit ans\"atze \cite{OMalley2016, kandala2017hardware, PhysRevX.8.031022, kandala2019error, Siddiqi2017, google2020hartree}. Regarding the latter, recent demonstrations \cite{stanisic2021observing} have used strategies outlined in \cite{JBL:JWFermiEnc}. These strategies utilize the Jordan-Wigner encoding to map $H_{FH}$ to qubit (Pauli) operators, initialize the qubits in a state $|\psi_0\rangle$ corresponding to the ground state of $T$ \cite{JBL:initialstate}, and then implement parameterized quantum circuits structured according to the Hamiltonian variational ansatz \cite{JBL:hva}. In Sec.~\ref{subsec:applyfhmod}, we formulate an FQA for the Fermi-Hubbard model that draws on these strategies.

Due to the constraints associated with current NISQ hardware, all of these demonstrations have been limited to relatively small problems, e.g., those involving tens of qubits and shallow quantum circuits. A critical challenge associated with scaling up VQE to larger problems will be the cost of the (nonconvex) classical optimization problem, which increases rapidly with the dimension of the parameter vector $\vec{\beta}$ and can quickly become intractable \cite{PhysRevLett.127.120502}. This is expected to make VQE prohibitively expensive for practical problems of interest. In this work, we consider FQAs that eliminate this computationally demanding classical optimization and instead operate by setting parameters according to a deterministic feedback law derived from the continuous-time theory of quantum Lyapunov control.

\subsection{Quantum Lyapunov Control}
\label{JBL:subsec:QLC}
Quantum Lyapunov control (QLC) is a framework for designing one or more time-dependent controls to drive a quantum dynamical system towards a desired objective in an asymptotic manner \cite{JBL:lyapunovsurvey, JBL:1272601}. To introduce this framework, we begin by considering a quantum system whose dynamics are dictated by the time-dependent Schr\"odinger equation (setting $\hbar = 1)$,
\begin{equation}
\label{JBL:system}
    i\frac{d}{dt}|\psi(t)\rangle = \big(H_p + H_d \beta(t)\big)|\psi(t)\rangle,
\end{equation}
where $H_p$ denotes the time-independent portion of the Hamiltonian; we refer to this as the \textit{problem} Hamiltonian, in direct analogy to the problem Hamiltonian in VQE. Meanwhile, $H_d$ denotes the control Hamiltonian that couples a time-dependent control function, $\beta (t)$, to the system; we refer to this as the \textit{driver} Hamiltonian, and note that the extension to cases with multiple control functions is straightforward, e.g., Ref.~\cite{JBL:longfalqon}. 

We consider the quantum control problem of designing $\beta (t)$ to drive $|\psi (t)\rangle$ to a state that minimizes an objective function $J$. QLC has been formulated for a variety of different objective functions (i.e., control Lyapunov functions), e.g., capturing the distance from a target state \cite{KUANG200898, 5415539, BEAUCHARD2007388}, the target state error \cite{MIRRAHIMI20051987}, and the expectation value of a target observable \cite{JBL:1272601, doi:10.1002/rnc.1748}. We restrict our attention to the observable case and define $J$ as the expectation value of $H_p$,
\begin{equation}
    J = \langle \psi(t) | H_p | \psi(t) \rangle.
    \label{JBL:Eq:expval}
\end{equation}
To solve this control problem, QLC seeks $\beta(t)$ such that $J$ decreases monotonically over time $t$. It is an asymptotic method, so the final time need not be specified \emph{a priori}. In particular, QLC seeks $\beta(t)$ that satisfies the derivative condition
\begin{equation}
\label{JBL:Eq:cond}
\frac{dJ}{dt} \leq 0, \forall t.
\end{equation}
Evaluating the left side of Eq.~\eqref{JBL:Eq:cond} using Eqs.~\eqref{JBL:system} and (\ref{JBL:Eq:expval}) yields
\begin{equation}
\begin{aligned}
    \frac{dJ}{dt} &= A(t)\beta(t),
    \label{JBL:eq:changeinexp}
\end{aligned}
\end{equation}
where we have introduced the abbreviated notation $A(t)$ to describe the time-dependent expectation value $A(t)\equiv \langle \psi(t) | i[H_d,H_p] | \psi(t) \rangle$. Given the form of Eq.~\eqref{JBL:eq:changeinexp}, the derivative condition in Eq.~\eqref{JBL:Eq:cond} can be satisfied for $\beta(t) = -w f(t,A(t))$, where $w > 0$ is a positive weight and $f(t,A(t))$ is a function selected such that $f(t,0)=0$ and $A(t)f(t,A(t))>0$ for all $A(t)\neq 0$ \cite{JBL:lyapunovsurvey}. The particular formulation utilized in the remainder of this work considers $w=1$ and $f(t,A(t)) = A(t)$, leading to the following control law 
\begin{equation}
\beta(t) = -A(t) 
\label{JBL:eq:feedbacklawchoice}
\end{equation}
that clearly satisfies the condition given by Eq.~\eqref{JBL:Eq:cond}, yielding $\frac{dJ}{dt} = -\big(A(t)\big)^2\leq 0$ for all $A(t)$. 

The convergence of QLC has been studied using the LaSalle invariance principle \cite{la1976stability}, leading to the identification of a set of sufficient conditions that, if satisfied, guarantee asymptotic convergence to the global minimum of Eq.~\eqref{JBL:Eq:expval} (see Appendix A of \cite{JBL:longfalqon} and \cite{JBL:1272601, doi:10.1002/rnc.1748}). However, these conditions are very stringent and are rarely satisfied in applications of interest. Nevertheless, convergence to the global minimum is often attained in practice, as evidenced by a growing body of numerical simulations \cite{JBL:1272601, KUANG2017164, JBL:falqon}, even in cases where the convergence criteria are violated. Better understanding of the necessary conditions for convergence of QLC, in order to close the gap between mathematical results and numerical observations, is a compelling open research problem. 

\subsection{FALQON and FQAs}\label{Sec:FQAs}

In this section, we develop the concept of FQAs and their connection to the continuous-time QLC framework outlined in Sec.~\ref{JBL:subsec:QLC}. This section is in line with the material in \cite{JBL:falqon}, which introduced FALQON as an FQA for solving combinatorial optimization problems. FALQON serves as the progenitor to the ground state preparation FQAs described in Sec.~\ref{JBL:sec:apply}, which utilize a similar alternating operator ansatz \cite{JBL:kremenetski2021quantum}. We begin by considering Eq.~\eqref{JBL:system}, whose general solution is given by
\begin{equation}
\label{JBL:timeordering}
|\psi (t) \rangle = \mathcal{T} e^{-i \int_0^t (H_p + H_d\beta(t'))dt'} |\psi (0) \rangle,
\end{equation}
where $\mathcal{T}$ denotes the time-ordering operator. FQAs are formulated by digitizing this evolution in two steps. First, time is discretized into a sequence of $\ell$ steps, and the Hamiltonian is approximated as time-independent over each step, such that
\begin{equation}
    |\psi(t+\Delta t)\rangle \approx e^{-i(H_p + \beta(t)H_d)\Delta t}|\psi(t)\rangle.
    \label{JBL:Eq:digitizedev}
\end{equation}
The second step is to approximate the evolution over each time step using Trotterization:
 \begin{equation}
    e^{-i(H_p + \beta_kH_d)\Delta t} \approx e^{-i\beta_kH_d\Delta t} e^{-iH_p\Delta t}.
    \label{JBL:Eq:trotterizedev}
\end{equation}
This sequence of approximations serves to break the complicated unitary governing the time evolution in Eq.~\eqref{JBL:timeordering} into a product of simpler exponentials, yielding a digitized formulation where
\begin{equation}
\begin{aligned}
    |\psi_\ell\rangle &=e^{-i\beta_\ell H_d\Delta t} e^{-iH_p\Delta t} \cdots e^{-i\beta_1H_d\Delta t} e^{-iH_p\Delta t}|\psi_0\rangle \\
    &=U_d(\beta_\ell)U_p...U_d(\beta_1)U_p|\psi_0\rangle.
    \label{JBL:Eq:falqev}
\end{aligned}
\end{equation}
In Eq.~\eqref{JBL:Eq:falqev}, we have adopted the notation $\beta_k \equiv \beta(k\Delta t)$; $|\psi_k \rangle \equiv |\psi(k\Delta t)\rangle$; $U_p \equiv e^{-iH_p\Delta t}$; and $U_d(\beta_k) \equiv e^{-i\beta_kH_d\Delta t}$, where $k=1,2,...,\ell$ indexes the digitized step, or layer, of the FQA. 

We now describe the mechanics of implementing FQAs. At the outset, an initial state $|\psi_0\rangle$ and a time step $\Delta t$ must be selected. As a practical matter, $|\psi_0\rangle$ should be chosen such that it is efficient to prepare on a quantum computer. In our studies, we select $|\psi_0\rangle$ to be the ground state of the driver Hamiltonian associated with the fermion/spin number subspace of interest \cite{JBL:projection}, and thus maintain a correspondence to quantum annealing. 

To run an FQA, a set of qubits are prepared in $|\psi_0\rangle$. Then, this state is evolved over a sequence of steps, using the quantum circuit structure in Eq. (\ref{JBL:Eq:falqev}). To set the values of the associated quantum circuit parameters $\beta_k$ at each step $k = 1,2,\cdots \ell$, FQAs utilize the feedback law 
\begin{equation}
\label{JBL:eq:feedbacklaw}
\beta_k = -A_{k-1},
\end{equation}
which is a digitized version of the control law given earlier in Eq.~\eqref{JBL:eq:feedbacklawchoice}, where
\begin{equation}
\label{eq:ak}
    A_k = \langle \psi_k | i[H_d,H_p] | \psi_k \rangle.
\end{equation}
We refer to Eq.~\eqref{JBL:eq:feedbacklaw} as a \emph{feedback law} because at each step, $A_{k-1}$ is ``fed back" to set the value of the next $\beta_k$. For sufficiently small $\Delta t$, this procedure guarantees that the cost function $J_k = \langle \psi_k |H_p|\psi_k\rangle$ will decrease monotonically as a function of layer $k$, i.e. $J_1\geq J_2\geq \cdots\geq J_\ell$, satisfying a digitized version of Eq.~\eqref{JBL:Eq:cond}. A bound for (positive) $\Delta t$ to ensure this property holds can be obtained according to the steps in Appendix A of \cite{https://doi.org/10.48550/arxiv.2301.04201}, and is given in terms of the spectral norms of $H_p$ and $H_d$ by 
\begin{equation}
\label{jbl:eq:timestepbound}
    \Delta t \leq \frac{1}{4||H_p||\,||H_d||^2}.
\end{equation}

\begin{figure}[htb]
\begin{center}
\scalebox{0.35}{\includegraphics{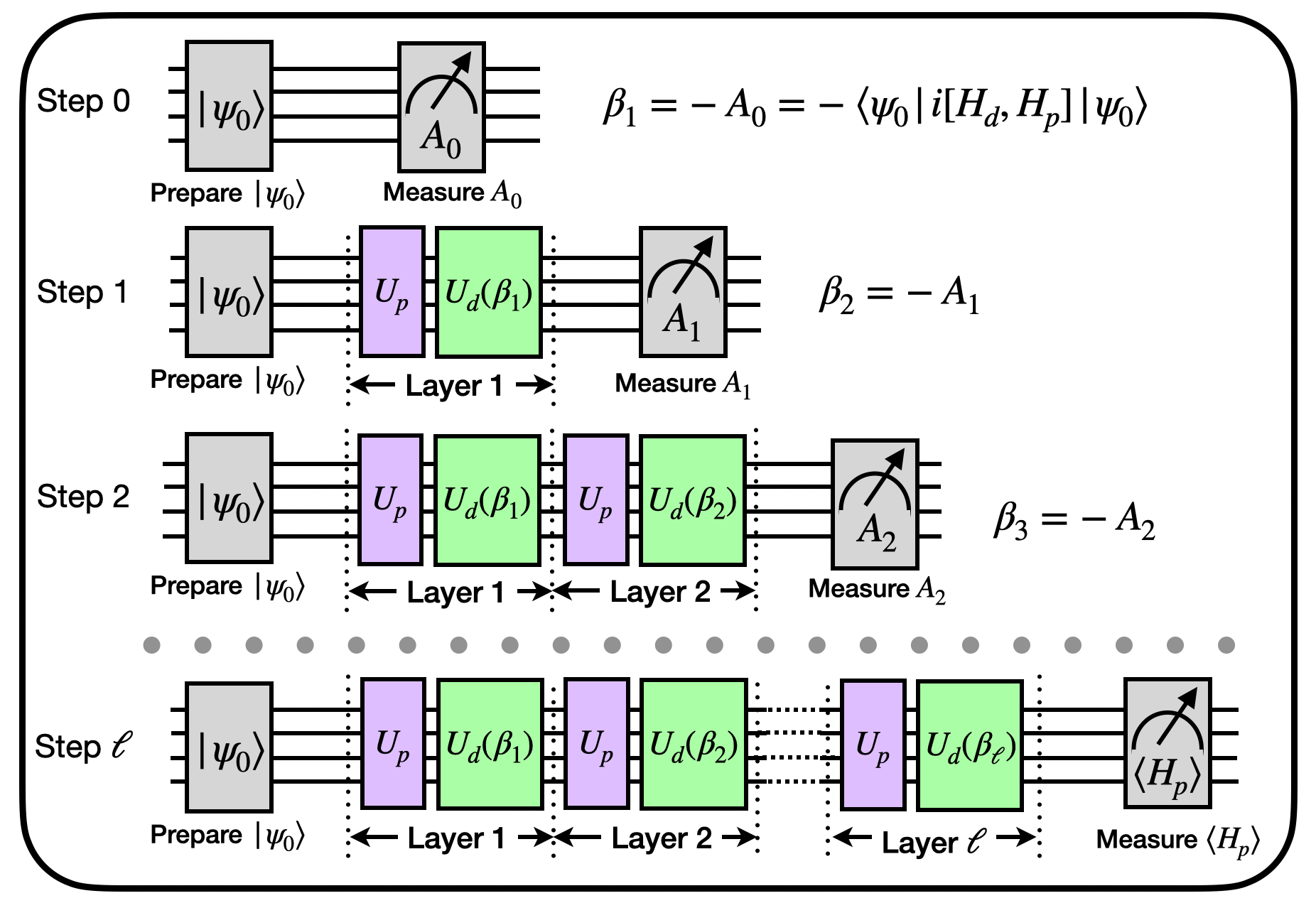}} \caption{The steps for implementing FQAs are shown (figure adapted from Ref.~\cite{JBL:falqon}). Each step $k$ involves extending the quantum circuit by one layer, concatenating the quantum circuit from the prior step $k-1$ with $U_d(\beta_k)U_p$. The value of $\beta_k$ at each layer $k$ is set using the results from the prior step $k-1$ according to the feedback law $\beta_k = -A_{k-1}$.}\label{fig2}
\end{center}\end{figure}

At each step $k$ of an FQA implementation, the quantum state $|\psi_k\rangle$ is prepared by applying the $k$-layer quantum circuit $U_k(\beta_k) U_p \ldots U_d(\beta_1) U_p$ to the (fixed) initial state $|\psi_0\rangle$. At the culmination of this circuit, the value of $A_k$ can be estimated, e.g., via repeated measurements of the observable $i[H_d,H_p]$, decomposed into a linear combination of Pauli operators, in the state $|\psi_k\rangle$. Finally, the estimate for $A_k$ is used to set the value of $\beta_{k+1}$ via the feedback law in Eq.~\eqref{JBL:eq:feedbacklaw}, and the FQA proceeds to step $k+1$, which follows the same procedure. We remark that each step $k$ corresponds to taking a step down in the direction of the layer-wise gradient $\frac{d}{d\beta_k}J$ with step size of $\Delta t$. For further implementation details we refer to Fig.~\ref{fig2} and \cite{JBL:falqon,JBL:longfalqon}.

A variety of other quantum ground state preparation algorithms are based on adaptive approaches where at each step, an additional layer is appended to the quantum circuit of the prior step, as in FQAs. A notable example is the Adaptive Derivative-Assembled Problem-Tailored Variational Quantum Eigensolver (ADAPT-VQE) \cite{grimsley2019adaptive} and its extension to combinatorial optimization applications \cite{PhysRevResearch.4.033029}. ADAPT-VQE operates by adaptively defining the structure of the quantum circuit in a layer-wise manner, in combination with classical optimization over the quantum circuit parameters. Other works have investigated the utility of layer-wise optimization of quantum circuit parameters for ans\"atze with a fixed structure, e.g., Refs.~\cite{carolan2020variational, skolik2021layerwise, PhysRevA.103.032607, campos2021training}. Optimization-free methods have also been explored, e.g., the quantum imaginary time evolution algorithm \cite{NatBrandao}, an algorithm based on Riemannian gradient flows \cite{https://doi.org/10.48550/arxiv.2202.06976}, and the randomized adaptive quantum state preparation algorithm \cite{https://doi.org/10.48550/arxiv.2301.04201}. Given this context, the distinguishing feature of FQAs is that they are derived from the continuous (real) time theory of QLC in a manner that produces problem-tailored quantum circuits with the structure given in Eq.~\eqref{JBL:Eq:falqev}, and parameterization defined in a layerwise manner according to the feedback law in Eq.~\eqref{JBL:eq:feedbacklaw}. 

\section{FQAs for Ground State Preparation}
\label{JBL:sec:apply}

\subsection{FQAs for the Fermi-Hubbard Model}
\label{subsec:applyfhmod}
In this section, we develop an FQA for preparing ground states of Fermi-Hubbard model. We begin by defining
\begin{equation}
H_p = H_{FH},\quad H_d = T,
\end{equation}
such that the feedback law from Eq.~\eqref{JBL:eq:feedbacklaw} is given by
\begin{equation}
\beta_{k} = -A_{k-1} =  -\langle \psi_{k-1}| i[T,V]|\psi_{k-1} \rangle.
\label{JBL:feedbacklaw}
\end{equation}
With these choices, an FQA can be implemented according to the steps discussed in Sec.~\ref{Sec:FQAs} and illustrated in Fig.~\ref{fig2}, using quantum circuits of the form given in Eq.~\eqref{JBL:Eq:falqev}.

\begin{figure}[htb]
\begin{center}
\scalebox{0.295}{\includegraphics{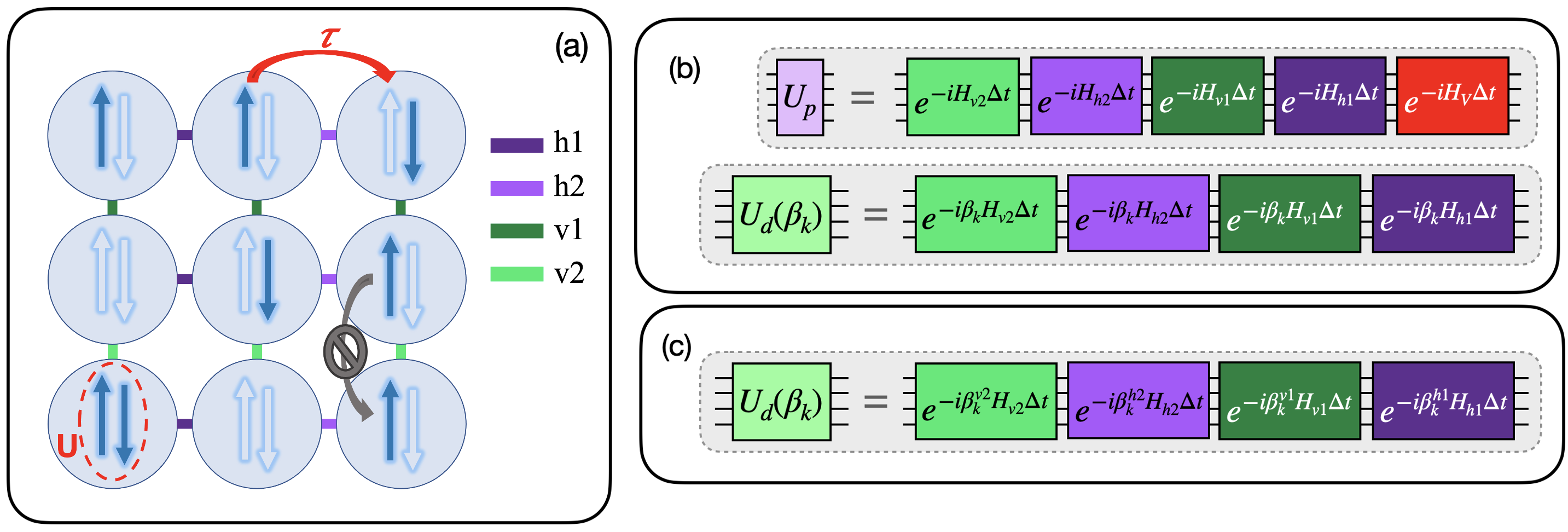}} \caption{In (a), a pictorial representation of 5 $\uparrow$ and 3 $\downarrow$ fermions on a $3 \times 3$ Fermi-Hubbard lattice is shown. Filled (dark blue) and unfilled (light blue) arrows on each site depict occupied and unoccupied spin orbitals, respectively. The repulsive on-site interactions between fermions have strength $U$, and the tunneling amplitude associated with fermions (of the same spin) hopping between adjacent lattice sites is denoted by $\tau$, as shown in red. Two fermions with the same spin cannot occupy the same state, and thus, these fermions are prohibited from hopping to adjacent states that are already occupied, as illustrated in gray. The green and purple bars notate the grouping of hopping terms used in the FQA formulation developed in Sec.~\ref{subsec:applyfhmod} in Eqs.~\eqref{JBL:eq:problemunitary} and \eqref{JBL:eq:oneparamhva}. Panels (b) and (c) show associated circuit diagrams, giving the specific decompositions of $U_p$ and $U_d$ from Fig.~\ref{fig2}, applied in the case of the Fermi-Hubbard model \cite{JBL:JWFermiEnc}. Panel (c) specifically shows the case for $U_d$ with multiple parameters, which is explored further in Appendix~\ref{JBL:appendixmcf}.
}
\label{fig:fh}
\end{center}\end{figure}

We now discuss a specific formulation that draws on the strategies from \cite{JBL:JWFermiEnc} and is used in the numerical simulations presented below in Sec.~\ref{JBL:Sec:results}. In particular, for $N_{sites}$ lattice sites arranged in a 1D or 2D square lattice, we utilize the Jordan-Wigner encoding to map the corresponding $2N_{sites}$ spin orbitals to $2N_{sites}$ qubits, such that each qubit encodes the occupation of an associated spin orbital. In this encoding, $T$ and $V$ map to Pauli operators as
\begin{equation}
    T = - \frac{\tau}{2} \sum_{\langle i,j\rangle} (X_iX_j + Y_iY_j)Z_{i+1}...Z_{j-1}
\end{equation}
and
\begin{equation}
    V = \frac{U}{4} \sum_{\langle a,b \rangle} (I-Z_a)(I-Z_b),
\end{equation}
respectively, where $i$ and $j$ index over qubits that encode spin orbitals on neighboring lattice sites associated with the same spin, and $a$ and $b$ index over qubits that encode spin orbitals on the same lattice site with opposite spin.

Using this encoding, we decompose the layers of the quantum circuits further, taking inspiration from the Hamiltonian variational ansatz \cite{JBL:hva}, as described in \cite{JBL:JWFermiEnc}. In particular, we separate $T$ into sets of mutually commuting terms, and then introduce an additional Trotter decomposition based on these sets. For a 1D model, $T$ is decomposed into two sets of mutually commuting terms: $H_{h1}$, composed of hopping terms associated with even-numbered sites, and $H_{h2}$ for odd sites. For a 2D model, $T$ has four sets of terms. The two additional sets of terms, $H_{v1}$ and $H_{v2}$, take into account the vertical dimension. This breakdown is depicted in Fig.~\ref{fig:fh}(a) (see legend). This leads to the decomposition 
\begin{equation}
    U_p = e^{-iV\Delta t}e^{-iH_{h1}\Delta t}e^{-iH_{v1}\Delta t}e^{-iH_{h2}\Delta t}e^{-iH_{v2}\Delta t}
    \label{JBL:eq:problemunitary}
\end{equation}
and
\begin{equation}
    U_d(\beta_k)= e^{-i\beta_kH_{h1}\Delta t}e^{-i\beta_kH_{v1}\Delta t}e^{-i\beta_kH_{h2}\Delta t}e^{-i\beta_kH_{v2}\Delta t},
\label{JBL:eq:oneparamhva}
\end{equation}
for $U_p$ and $U_d(\beta_k)$, respectively, as depicted in Fig.~\ref{fig:fh}(b). We note that this framework can be adapted to contain multiple parameters per layer in the decomposition for $U_d$, as depicted in Fig.~\ref{fig:fh}(c) and discussed further in Appendix~\ref{JBL:appendixmcf}. Finally, in keeping with Ref.~\cite{JBL:JWFermiEnc}, we consider a fixed initial state $|\psi_0\rangle$ that corresponds to the ground state of $T$. This state can be prepared via Givens rotations \cite{JBL:initialstate}. 

A key consideration for this FQA, especially for NISQ device implementations, will be the sampling cost associated with estimating the expectation value $A_k$ at each layer $k$ to evaluate the feedback law in Eq.~\eqref{JBL:feedbacklaw}. In the Jordan-Wigner encoding, this expectation value is given by $A_k = \langle\psi_k|i[T,V]|\psi_k\rangle$, where
\begin{equation}
    [T,V]= -\frac{i\tau U}{4}\sum_{i,j} (X_iY_jZ_a - X_jY_iZ_a - X_iY_jZ_b + X_jY_iZ_b)Z_{i+1}...Z_{j-1},
    \label{eq:JWencodedcomm}
\end{equation}
with $a$ and $b$ indexing the orbitals of opposite spin. Using parallelization techniques similar to those in \cite{gokhale2019minimizing, Verteletskyi2020, reggio2023fast, anastasiou2023really, zhu2023optimizing}, i.e., that are based on the grouping of mutually commuting Pauli strings in Eq. (\ref{eq:JWencodedcomm}), we find that the number of measurement samples needed for estimating the expectation value of each $A_k$ is given by
\begin{equation}
    N_{samples} = (4n_c+6)m
\end{equation}
for an $n_r\times n_c$ lattice, where $n_c$ is the horizontal dimension of the Fermi-Hubbard lattice (i.e., number of columns), and $m$ is the number of measurement samples required for estimating the expectation value of individual Pauli strings to a desired precision. Full details can be found in Appendix~\ref{app:fhmod}. 

\subsection{FQAs for Molecular Hamiltonians}
\label{sec:mhfqa}
We now develop an FQA for preparing ground states of second-quantized molecular Hamiltonians as introduced in Sec.~\ref{sec:molhamint}. Similar to our approach for the Fermi-Hubbard model, we define
\begin{equation}
    H_p = H_{MH}, \quad H_d = H_1,
\end{equation}
such that the feedback law from Eq.~(\ref{JBL:eq:feedbacklaw}) is given by
\begin{equation}
\label{eq:mhfeedbacklaw}
    \beta_k = -A_{k-1} = -\langle \psi_{k-1} | i [H_1, H_2] | \psi_{k-1} \rangle.
\end{equation}
With these design choices, an FQA can be implemented using the alternating operator ansatz described in Sec.~\ref{Sec:FQAs}. One layer of the circuit for this FQA is given by 
\begin{equation}
    U_d(\beta_k) U_p = e^{-i\beta_k H_1 \Delta t} e^{-iH_{MH} \Delta t}.
\end{equation}
Just as with the Fermi-Hubbard model, we utilize the Jordan-Wigner encoding to map the spin orbitals to qubits and separate $H_1$ and $H_{MH}$ into sets of mutually commuting terms, using an additional Trotter decomposition based on these sets.

The commutator in Eq.~(\ref{eq:mhfeedbacklaw}) can be expressed in terms of fermionic creation and annihilation operators by
\begin{equation}
\label{mhfermopcomm}
[H_1, H_2] = \frac{1}{2} \sum_{ij} t_{ij} \sum_{abcd} u_{abcd} [(a_i^\dagger a_b^\dagger \delta_{ja} + a_a^\dagger a_i^\dagger \delta_{jb})a_c a_d + a_a^\dagger a_b^\dagger (a_j a_d \delta_{ic} + a_c a_j \delta_{id})].
\end{equation}
The explicit Pauli decomposition of Eq.~\eqref{mhfermopcomm} via Jordan-Wigner is not written out here for space considerations, but the various terms in this decomposition are described in Appendix~\ref{app:molham}. As before, we can reduce the sampling cost of estimating each $A_k$ by parallelizing measurements by grouping mutually commuting Pauli strings. To this end, we find that the number of measurement samples needed for estimating the expectation value of each $A_k$ in the molecular setting is given by
\begin{equation}
\begin{aligned}
N_{samples} &= \left(1 + 2^2 \binom{n}{2} + 2^4 \binom{n}{4}\right)m = \left( \frac{2}{3}n^4 - 4n^3 + \frac{28}{3}n^2-6n+1 \right)m,
\end{aligned}
\end{equation}
where $n$ is the number of qubits (equivalently, spin-orbitals). For further details, see Appendix~\ref{app:molham}.

\section{Numerical Illustrations}
\subsection{Fermi-Hubbard}
\label{JBL:Sec:results}

In this section, we explore the performance of the FQA for ground state preparation of the Fermi-Hubbard model through numerical simulations. For these demonstrations, we look specifically at open boundary conditions. 

\begin{figure}[ht]
\begin{center}
\includegraphics[scale=.305]{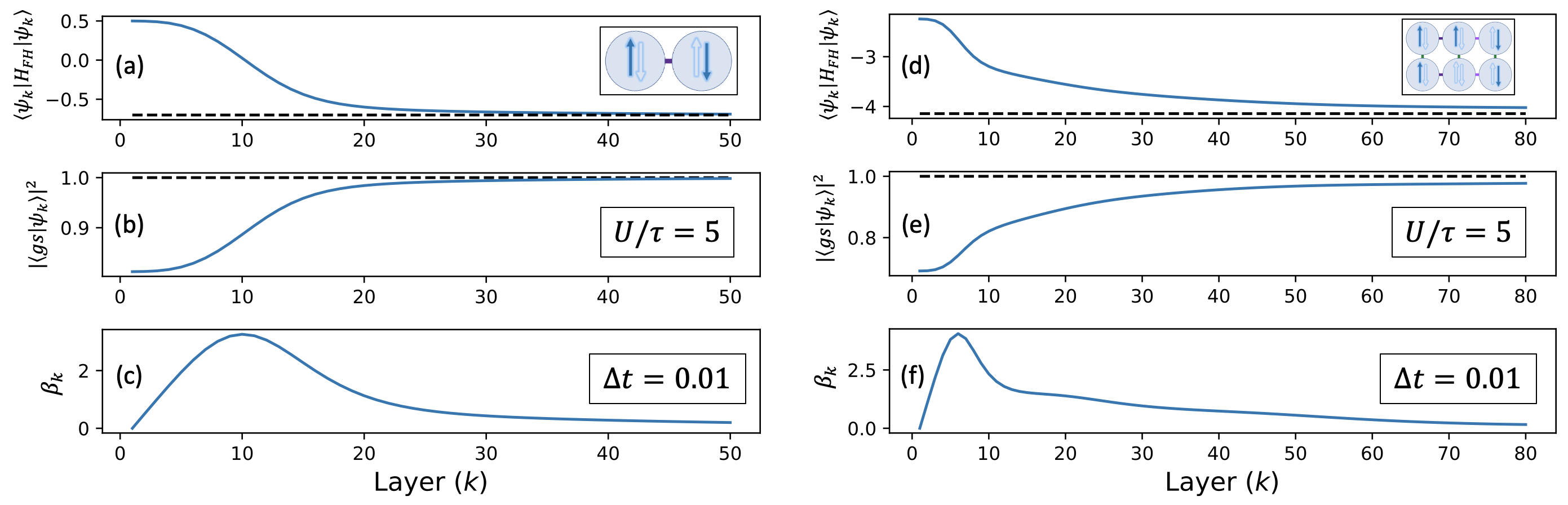}
\caption{Numerical simulations of the FQA formulated in Sec.~\ref{JBL:sec:apply} applied to two different Fermi-Hubbard lattices: a $1\times 2$ lattice with 1$\uparrow$ and 1$\downarrow$ fermion in (a)-(c) and a $2\times 3$ lattice with 3$\uparrow$ and 2$\downarrow$ fermions in (d)-(f). Simulations use $\Delta t = 0.01$, and $U/\tau=5$. In panels (a) and (d), the dashed lines correspond to the true ground state energy, i.e., corresponding to $J_{min} = \langle gs|H_{FH}|gs\rangle$, where $|gs\rangle$ denotes the ground state of the Fermi-Hubbard Hamiltonian $H_{FH}$ associated with the appropriate fermion/spin sector. The blue curves show the convergence of the cost function, $J = \langle\psi_k|H_{FH}|\psi_k\rangle$ as a function of layer $k$, when qubits are initialized in the ground state of $T$. Panels (b) and (e) show like results for the squared overlap $|\langle gs|\psi_k\rangle|^2$, and panels (c) and (f) show the quantum circuit parameters obtained via the feedback law $\beta_k = -A_{k-1}$ according to the steps in Fig.~\ref{fig2}.}
\label{fig:fhres}
\end{center}\end{figure}

We begin by considering the simplest instance of the Fermi-Hubbard model: a two-site lattice at half-filling. The results are presented in Fig.~\ref{fig:fhres}(a-c) for $\Delta t = 0.01$ and $U/\tau = 5$. Using the same choice of $\Delta t$ and $U/\tau$, we present analogous results for a $2 \times 3$ lattice with just under half-filling in Fig.~\ref{fig:fhres}(d-f). Panels (a) and (d) in Fig.~\ref{fig:fhres} illustrate that the value of the objective function decreases monotonically with respect to layer, as expected given our choice of feedback law from Eq.~\eqref{JBL:feedbacklaw}. Panels (b) and (e) show that as the value of the objective function decreases, the squared overlap with the ground state, $|\langle gs|\psi\rangle|^2$, approaches 1, behavior that should be satisfied by any successful ground state preparation algorithm. Panels (c) and (f) show the $\beta$ parameter values used in the circuit, as obtained by the feedback law in Eq.~\eqref{JBL:feedbacklaw}.

\begin{figure}[htb]
\begin{center}
\scalebox{0.3}{\includegraphics{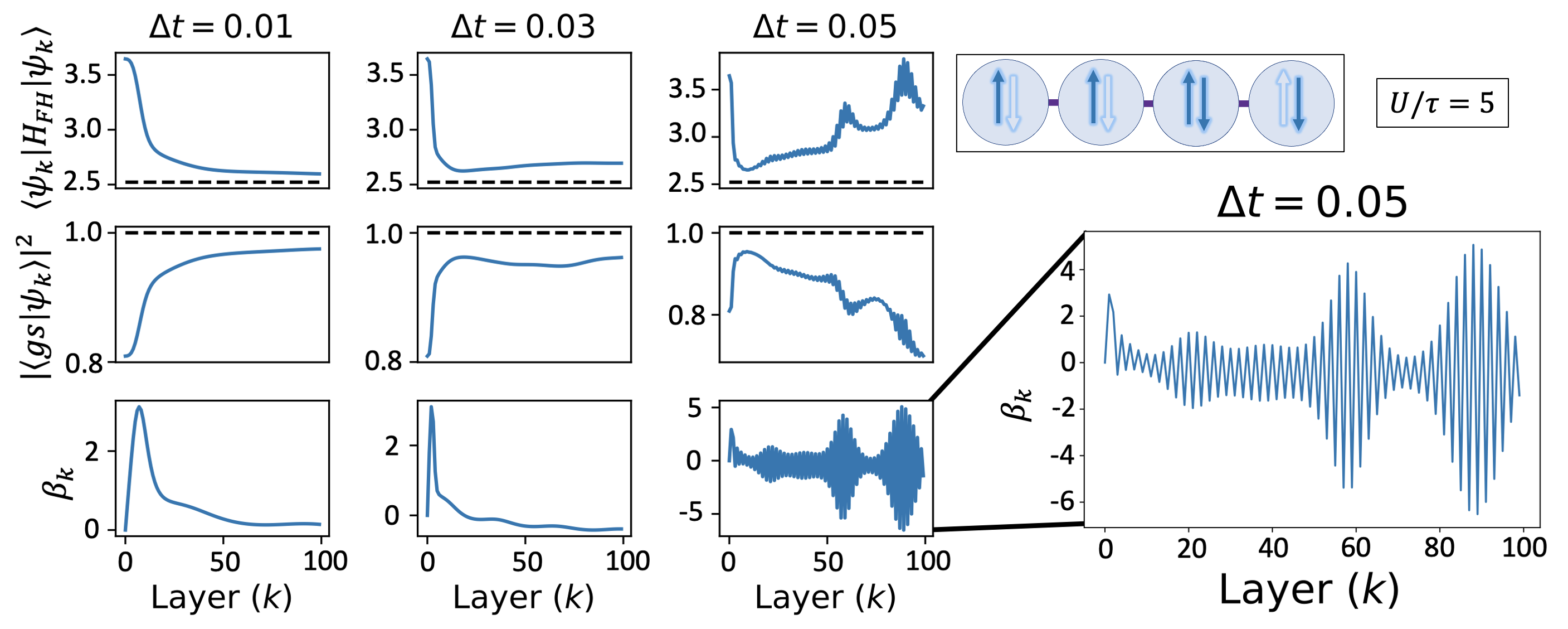}} \caption{Numerical simulations of the FQA formulated in Sec.~\ref{JBL:sec:apply} applied to a $1\times 4$ Fermi-Hubbard lattice with $3\uparrow 2\downarrow$ fermions for $U/\tau = 5$. Results are presented for $\Delta t = 0.01, 0.03$, and 0.05, and illustrate the impact that varying $\Delta t$ can have on FQA performance.}\label{JBL:fig:timestepdiverge}
\end{center}\end{figure}

In addition to the choice of feedback law for determining the $\beta_k$ parameters, this method also requires a choice for $\Delta t$. In most of the simulations that we performed, we found that $\Delta t=0.01$ was a suitable choice, as shown in Fig.~\ref{JBL:fig:timestepdiverge}. This indicates that good performance can be obtained in practice for values of $\Delta t$ that significantly exceed the bound presented in Eq.~(\ref{jbl:eq:timestepbound}). When $\Delta t$ is too large, however, we find the algorithm fails to converge. In this regime, the $\beta_k$ parameters oscillate dramatically and the objective function fails to decrease monotonically. In practice, FQAs do not require objective function evaluations at each step, only the estimation of $A_k$ to set each $\beta_k$ value. As such, signs of oscillatory behavior in $\beta_k$ (or equivalently, in the measurement record of $A_k$) can serve as an indication that $\Delta t$ is too large and the algorithm is not converging.

\begin{figure}[htb]
\begin{center}
\includegraphics[scale=.305]{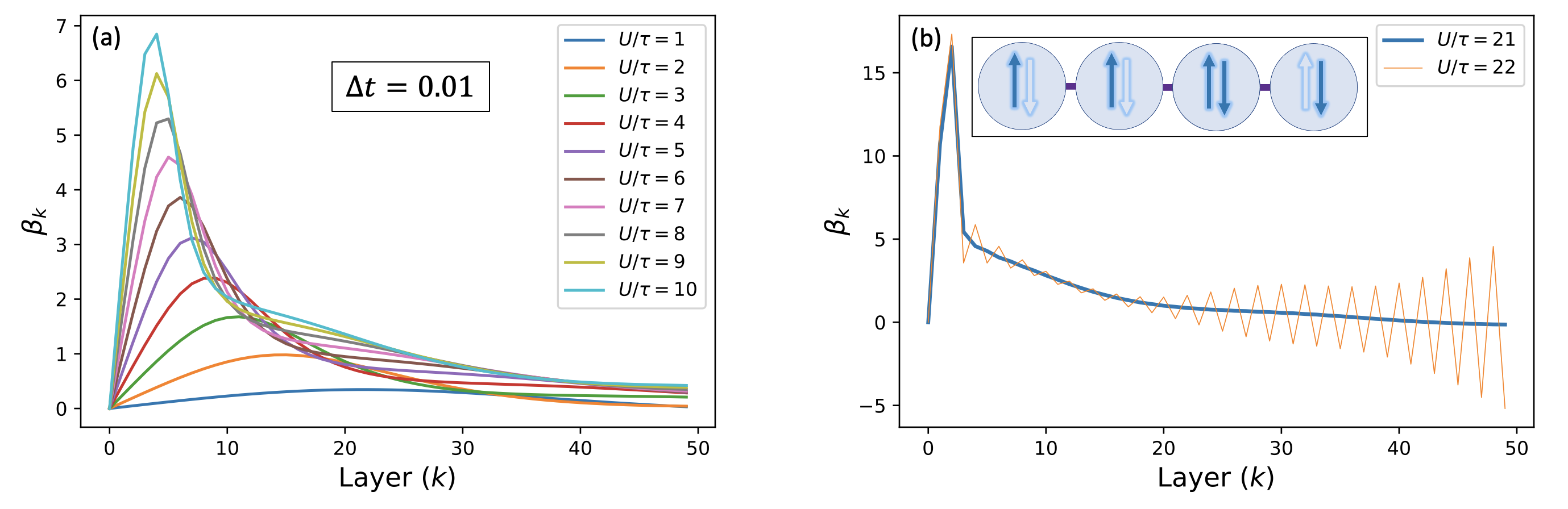}
\caption{Numerical simulations of the FQA formulated in Sec.~\ref{JBL:sec:apply} applied to a $1\times 4$ Fermi-Hubbard lattice with $3\uparrow 2\downarrow$ fermions and $\Delta t = 0.01$. Panel (a) shows $\beta_k$ for a variety of $U/\tau$ values. Panel (b) shows $\beta_k$ for $U/\tau = 21$ and $U/\tau = 22$, and the oscillations present in $\beta_k$ for the latter indicate that a reduction in $\Delta t$ is needed. }\label{figure6}
\end{center}\end{figure}

Fig.~\ref{figure6} explores how $\beta_k$ parameter curves vary for different regimes of $U/\tau$. Recall that the initial ground state corresponds to the ground state of $T$ (our choice of driver). Increasing $U/\tau$ then puts more relative weight to the portion of the Hamiltonian not captured with this choice of initial state. To compensate for this, intuition suggests that a sharper increase in the $\beta_k$ parameters should take place. This intuition is matched by the numerical results in Fig.~\ref{figure6}(a).
As $U/\tau$ increases, so do the $\beta_k$ parameters. We also observe an oscillatory behavior in Fig.~\ref{figure6}(b) that indicates a reduction in $\Delta t$ is needed. This is not surprising given the form of Eq.~(\ref{jbl:eq:timestepbound}), i.e., given that the bound for $\Delta t$ scales inversely with $||H_p||$. Increasing $U/\tau$ in Fig.~\ref{figure6} increases $||H_p||$, and should thus be coupled with an associated decrease in $\Delta t$.

\begin{figure}[htb]
\begin{center}
\scalebox{0.26}{\includegraphics{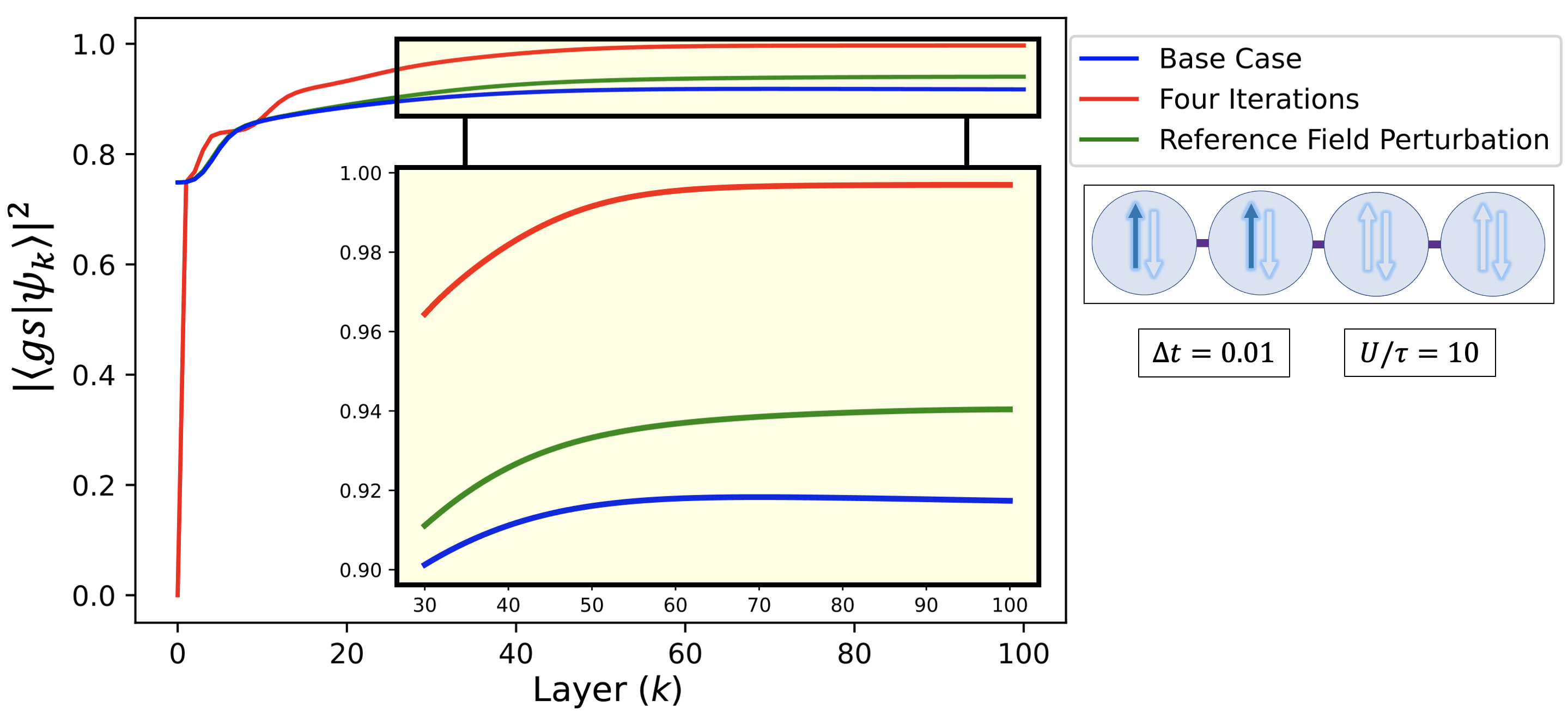}} \caption{Numerical simulation results for a $1\times 4$ Fermi-Hubbard lattice with $1\uparrow 1\downarrow$ fermions for $\Delta t=0.01$, in the $U/\tau=10$ regime. The blue curves show the convergence of the squared overlap $|\langle gs|\psi_k\rangle|^2$ for the the base case, i.e., the FQA formulated in Sec.~\ref{JBL:sec:apply} with no modifications. The green curves show the new convergence when a reference field perturbation is added to $\beta_k$. The red curves show the new convergence using an iterative procedure after four iterations. We observe that the iterative modification provides the most significant improvement for this problem instance, although the reference field modification also leads to improvements over the base case.}\label{JBL:fig:improv}
\end{center}\end{figure}

Fig.~\ref{JBL:fig:improv} presents results from implementing two modified versions of the FQA aimed at improving convergence, following Refs.~\cite{JBL:falqon,JBL:longfalqon}. These modifications include an iterative method and a reference field perturbation method (see Sec III B and III A in Ref.~\cite{JBL:longfalqon}, respectively, noting the correction \bibnote[bibnotecorrection]{The reference field perturbation heuristic is motivated in Sec. IIIA of \cite{JBL:longfalqon}, which discusses the continuous-time formulation. The first line of Eq. (14) in this section should read $\frac{d}{dt}E_{p,(b)} = \langle\psi(t)|i [H(t), H_{p,(b)}(t)] |\psi(t)\rangle + \frac{d\lambda(t)}{ dt} \langle\psi(t)|H_d |\psi(t)\rangle$, with the following lines of the equation updated accordingly. Consequently, the subsequent reasoning in Sec. IIIA of \cite{JBL:longfalqon} holds only if $\Big|\frac{d\lambda(t)}{ dt} \langle\psi(t)|H_d |\psi(t)\rangle\Big|$ is negligible, e.g., if the variation in the reference field perturbation, $\lambda(t)$, with time is sufficiently slow.}). For the former, the FQA is simulated for $\ell$ layers using the feedback law in Eq.~\eqref{JBL:feedbacklaw}. After acquiring the $\beta_k$ parameters for each layer, a second iteration is performed, which involves adding the original $\beta_k$ values to the newly obtained parameter values. This procedure is then repeated for a desired number of additional iterations. The results after four iterations are shown in red in Fig.~\ref{JBL:fig:improv}. We also explore adding a slowly-varying, reference field perturbation to the calculated $\beta_k$ parameters, resulting in new parameters $\beta_k'$. The perturbation is selected to be a linear ramp with negative slope, such as $\beta_k' = \beta_k + 0.5 (1 - k/\ell)$), with results shown in green. Both modifications result in improvements to convergence. We note that improving convergence may also be possible through the addition of random kicks in $\beta_k$, per the Appendix of Ref.~\cite{JBL:falqon}. Convergence to the ground state, even from arbitrary (random) initial states, could also be enabled through combination with randomized adaptive quantum state preparation, following Ref.~\cite{https://doi.org/10.48550/arxiv.2301.04201}.

\begin{figure}[htb]
\begin{center}
\scalebox{0.305}{\includegraphics{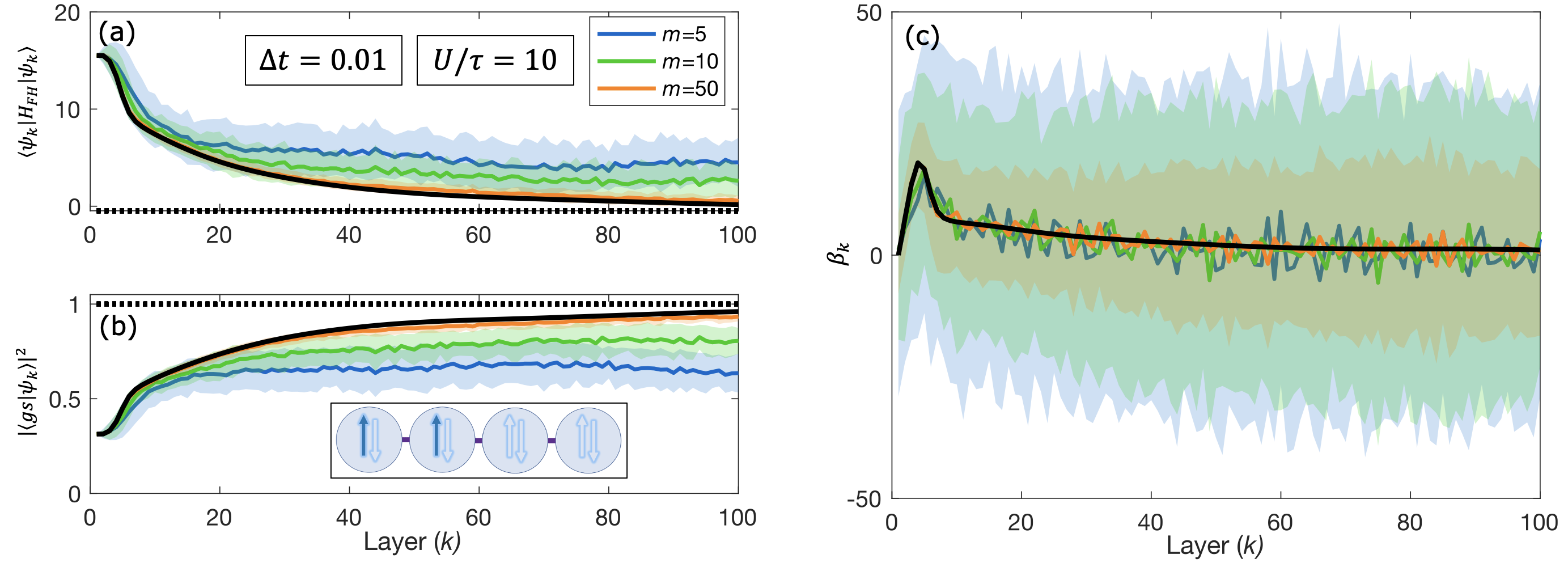}} \caption{Simulation results investigating the impact of sampling noise on FQA performance for a $1\times 4$ instance of the Fermi-Hubbard model with $2\uparrow 2\downarrow$ fermions, $U/\tau = 20$, and $\Delta t = 0.005$. Panel (a) shows $\langle\psi_k|H_{FH}|\psi_k\rangle$ plotted as a function of layer, $k$. The solid black curves show the ideal, noiseless  case (i.e., $m=\infty$) for comparison. The blue, green, and orange solid curves show the mean taken over 100 realizations of a sampling process where $m=5,10,50$ measurement samples are used to evaluate $A_k$, $k=1,\cdots,100$, respectively. The shading of corresponding color shows the associated standard deviation over these realizations. Panels (b) and (c) present analagous results for $|\langle gs|\psi_k\rangle|^2$ and $\beta_k$. }
\label{JBL:fig:robust}
\end{center}\end{figure}

We conclude this section by investigating the practical impact of sampling noise on the performance of the FQA for Fermi-Hubbard ground state preparation, with results presented in Fig.~\ref{JBL:fig:robust}. Specifically, we consider the situation where a finite number of measurement samples are utilized to estimate each value of $A_k$. Finite sampling will lead to deviations between the estimate for $A_k$ and the true, ideal expectation value. These noisy estimates of $A_k$ will then impact the algorithm directly through the feedback law for $\beta_k$. To simulate this, we sample measurement outcomes at each layer $k$ from the multinomial probability distribution, defined by the state $|\psi_k\rangle$, over the set of associated eigenvalues of $i[H_d,H_p]$. The results in Fig.~\ref{JBL:fig:robust}(a) and (b) show that the FQA has good performance even in the presence of significant sampling noise, as shown in Fig.~\ref{JBL:fig:robust}(c), i.e., with the results for $m=50$ (orange) tracking closely with the ideal results, i.e., for $m=\infty$ (black). This robustness can be understood in the context of the flexibility in defining a feedback law for setting $\beta_k$ to satisfy Eq. (\ref{JBL:Eq:expval}), i.e., we selected $\beta_k = -A_{k-1}$, but even noisy estimates of $A_{k-1}$ can lead to the satisfaction of Eq. (\ref{JBL:Eq:expval}) and good performance. 

\subsection{Molecular Hamiltonians}
\label{sec:molham}
In this section, we generalize the presented results to models in chemistry beyond the Fermi-Hubbard model, examining performance of the FQA for second-quantized molecular Hamiltonians introduced in Sec.~\ref{sec:mhfqa}. We look specifically at preparing the ground states of the hydrogen molecule (H$_2$), lithium hydride (LiH), and water (H$_2$O). We use OpenFermion-PySCF \cite{JBL:openfermion} to calculate the one- and two-electron integrals for these molecules in the STO-3G basis.
\begin{figure}[htb]
\begin{center}
\includegraphics[scale=.513]{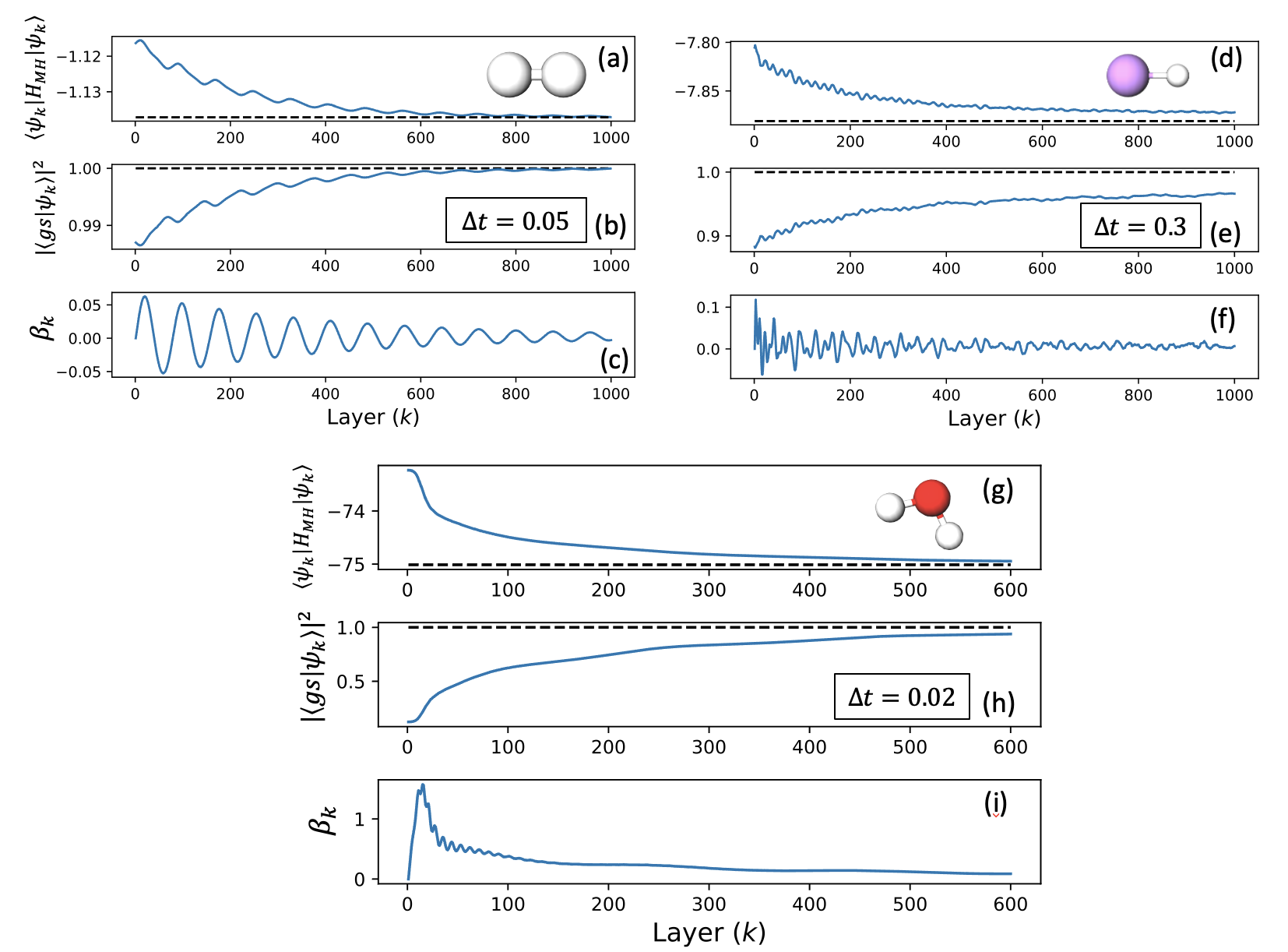} 
\caption{Demonstration of FQA for ground state preparation for three different molecular Hamiltonians. Panels (a-c) show simulations for the hydrogen molecule, (d-f) show simulations for lithium hydride, and (g-i) show simulations for water. Panels (a, d, g) show the value of the objective function, $\langle \psi_k | H_{MH} | \psi_k \rangle$, at each layer of the FQA, (b, e, h) show the squared overlap with the ground state, $|\langle gs | \psi_k \rangle|^2$, and (c, f, i) show the parameter values found via the feedback law in Eq. (\ref{eq:mhfeedbacklaw}).}
\label{fig:mhres}
\end{center}\end{figure}
Following the same convention as Fig.~\ref{fig:fhres}, Fig.~\ref{fig:mhres}(a,d,g) illustrate the monotonic decrease in the value of the objective function with respect to layer. The middle panels show that the squared overlap with the ground state, $|\langle gs|\psi\rangle|^2$, approaches 1 as the objective function decreases. The bottom panels show the $\beta$ parameter values used in the circuit, as obtained by the feedback law in Eq.~\eqref{eq:mhfeedbacklaw}.

Fig.~\ref{fig:mhres}(a-c) shows results for the hydrogen molecule with bond length 0.7474 \AA, using four spin orbitals (qubits). The oscillatory behavior seen in the $\beta$ parameter curve for this instance differs qualitatively from previous oscillatory behavior observed in Fig.~\ref{JBL:fig:timestepdiverge}, where Fig.~\ref{fig:mhres}(c) more smoothly traces out the oscillations. This sinusoidal shape is not due to our choice of timestep, but is instead an artifact of our initial state being a superposition of just two states, i.e., the ground state and one of the excited states. As a result, the parameter curve oscillates with a frequency proportional to the energy gap between these two states. The feedback law from Lyapunov control finds this resonant frequency in order to induce the transition to the lower energy state.

In Fig.~\ref{fig:mhres}(d-f), we show results for lithium hydride with a bond length of 1.45 \AA, using 12 spin orbitals. This molecule required somewhat deeper circuits for convergence, relative to other instances we explored. Fig.~\ref{fig:mhres}(g-i) shows results for water with 14 spin orbitals. The bond lengths between the hydrogen and oxygen atoms are set to 0.958 \AA, with an angle between these bonds of 104.5\textdegree. The resulting parameter curve for water in Fig.~\ref{fig:mhres}(i) takes a shape more similar to those observed for instances of the Fermi-Hubbard model.

\section{Discussion: FQAs, Adiabatic Ground State Preparation, and VQE}
\label{Sec:Discussion}

In this section, we contextualize the FQA developed here relative to alternative strategies based on analogous quantum circuit structures. A first example is adiabatic ground state preparation \cite{doi:10.1126/science.1113479}, which has strong ties to adiabatic computation \cite{https://doi.org/10.48550/arxiv.quant-ph/0001106, doi:10.1126/science.1057726, PhysRevA.81.032308}. Adiabatic ground state preparation is conventionally an optimization-free approach, similar to FQAs. It proceeds by initializing a quantum system in the ground state of a Hamiltonian, $H_d$, that is relatively simple to prepare, and then evolving this state under the time-dependent Hamiltonian 
\begin{equation}
    H(t) = w(t)H_p + u(t) H_d
    \label{eq:anneal}
\end{equation}
for a total time $T$, where $w(t) = 1-u(t)$ and $u(t)$ denotes the annealing schedule with $u(0)=1$ and $u(T) = 0$, such that $H(t)$ interpolates between $H_d$ and $H_p$ over the course of the time evolution. For sufficiently large $T$ and sufficiently slow evolution, the state of the quantum system will remain in the instantaneous ground state of $H(t)$, such that the system converges adiabatically to the ground state of the target Hamiltonian $H(T)=H_p$ at the final time, per the adiabatic theorem \cite{adiabatictheorem}. Certain formulations of adiabatic quantum state preparation only parameterize one term in $H(t)$, i.e., setting $w(t) = 1$ or setting $u(t) = 1$ for all $t$, where in the latter case, $w(t)$ is the annealing schedule with $w(0)=0$ and $w(T) = 1$.

To implement adiabatic ground state preparation on a quantum computer, time evolution under $H(t)$ in Eq.~\eqref{eq:anneal} can be simulated digitally using a variety of candidate Hamiltonian simulation approaches, including first-order Trotterization, i.e., according to the steps outlined in Sec.~\ref{Sec:FQAs} in Eqs.~\eqref{JBL:timeordering}-\eqref{JBL:Eq:trotterizedev}, leading to quantum circuits of the form
\begin{equation}
\begin{aligned}
    |\psi_\ell\rangle &=e^{-iu_\ell H_d\Delta t} e^{-iw_\ell H_p\Delta t} \cdots e^{-iu_1H_d\Delta t} e^{-iw_1H_p\Delta t}|\psi_0\rangle \\
    &=U_d(u_\ell)U_p(w_\ell)...U_d(u_1)U_p(w_1)|\psi_0\rangle,
    \label{annealcircuits}
\end{aligned}
\end{equation}
where $\ell$ denotes the number of discrete time steps of size $\Delta t$ used to simulate the time evolution of the system for $t\in[0,T]$, with $u(k\Delta t)$ and $w(k\Delta t)$ denoted using the shorthand $u_k$ and $w_k$. We observe that the quantum circuit in Eq.~\eqref{annealcircuits} has the same structure as the quantum circuit used in FQAs, per Eq.~\eqref{JBL:Eq:falqev}. This is because Eqs.~\eqref{annealcircuits} and \eqref{JBL:Eq:falqev} are both derived in a similar manner, i.e., beginning with time-dependent Hamiltonians given by Eq.~\eqref{eq:anneal} and 
\begin{equation}
H(t) = H_p+\beta(t) H_d, 
\label{eq:fqahamiltonian}
\end{equation}
respectively, which share a common structure when $w(t) = 1$ $\forall t$.

The key difference between adiabatic ground state preparation and FQAs then lies in how the elements $u(t)$ and $\beta(t)$ are defined. In adiabatic ground state preparation, the total time $T$ and the complete annealing schedule $u(t)$, $t \in [0,T]$, must both be specified \emph{a priori}. A common choice of annealing schedule is a ramp that interpolates linearly between $u(0) = 1$ and $u(T) = 0$. However, many other possibilities have been considered, and determining optimal annealing schedules is a difficult task in general. In contrast, the total time $T$ (or in the digitized picture, the number of total layers $\ell$) for FQAs need not be specified \emph{a priori}. That is, FQAs can continue stepping forward according to the procedure outlined in Sec.~\ref{Sec:FQAs} and diagrammed in Fig.~\ref{fig2}, i.e., at each step incrementing $k$ by 1 and appending an additional layer onto the quantum circuit, until desired convergence criteria are satisfied (e.g., $|A_k|\leq \epsilon$). In addition, the values of each $\beta_k$ are not defined upfront for FQAs, but are determined sequentially at each step as the FQA runs, according to the feedback law in Eq.~\eqref{JBL:eq:feedbacklaw}. We anticipate that the use of feedback in this manner could have benefits for FQAs with respect to performance, but collecting this feedback also requires additional resources, such as measurements to estimate $A_k$ at each later $k$, that are not needed in adiabatic quantum state preparation. A numerical comparison between the performance of these two strategies, specifically for solving the MaxCut problem, can be found in Ref.~\cite{JBL:longfalqon}. However, a careful exploration of the cost/benefit tradeoffs associated with FQAs, and relating the capabilities of FQAs to adiabatic quantum state preparation, would constitute valuable future research. 

We now turn to VQE, as outlined in Sec.~\ref{sec:vqe}, as a second example of a strategy that can utilize analogous quantum circuit structures. Research exploring applications of VQE to the Fermi-Hubbard model has often considered quantum circuits structured according to the Hamiltonian variational ansatz \cite{JBL:hva}, which is a layered ansatz. The formulation in \cite{JBL:JWFermiEnc} involves layers formed by a sequence of five parameterized unitaries generated by $V,H_{h1},H_{h2},H_{v1},H_{v2}$, as discussed in Sec.~\ref{JBL:sec:apply}. The variational parameters associated with each of these unitaries are classically optimized over in order to obtain a parameter configuration that minimizes the cost function $J = \langle\psi|H_{FH}|\psi\rangle$. The quantum circuits used for the FQA formulated in Sec.~\ref{JBL:sec:apply} are deliberately structured similarly, i.e., involving products of the same unitary operations. We note however that in the FQA formulation, a product of nine exponentials form each layer $k$, with the decompositions of $U_p$ and $U_d(\beta_k)$ given in Eqs.~\eqref{JBL:eq:problemunitary} and (\ref{JBL:eq:oneparamhva}), respectively. 

The primary difference between VQE and FQAs is that VQE operates via a classical optimization loop, while FQAs do not (see Fig.~\ref{figure1}). The classical optimization in VQE is expected to become a computational bottleneck as the number of variational parameters present in the quantum circuit increases \cite{PhysRevLett.127.120502} due to the difficulty of searching for the optimal parameter configuration in a high-dimensional parameter space. The number of variational parameters typically correlates to the depth of the associated ansatz, and as quantum hardware improves, we expect a push toward VQE with deeper circuits that will make the challenge associated with the classical optimization increasingly relevant. However, there is a tradeoff here: if the optimal parameter configuration can be found, classical optimization enables VQE to prepare ground states using relatively shallow circuits, making it an appealing option for implementation on near-term quantum hardware. Meanwhile, in FQAs the quantum circuit parameters are assigned values according to the feedback law in Eq.~\eqref{JBL:eq:feedbacklaw}. Consequently, FQAs leverage deeper circuits with high parameter counts, without suffering from the computational bottleneck associated with classically searching the associated parameter space. These tradeoffs between circuit depth and classical optimization cost suggest that VQE and FQAs have different regimes of applicability: VQE is applicable in the presence of limited quantum resources (circuit depth), the ability to estimate $J$ at each optimization iteration via qubit measurements, and sufficient classical resources to perform the needed optimization, while FQAs are applicable in the presence of sufficient quantum resources, combined with the ability to estimate $A_k$ via qubit measurements at each step to evaluate the feedback law in Eq.~\eqref{JBL:eq:feedbacklaw}. We conclude this discussion by noting that many theoretically promising approaches for ground state preparation likewise appear to require deep quantum circuits \cite{reiher2017elucidating, lemieux2021resource, pathak2023quantifying, gratsea2022reject}, i.e., beyond the capabilities of current quantum computers and thus requiring fault-tolerant quantum computation. Many questions regarding how FQAs fit in relative to these other ground state preparation algorithms remain open; performing an accounting of the different resources required by each and clarifying the tradeoffs  between them would be helpful next steps in answering them. 

\section{Summary and Future Avenues}
\label{JBL:Sec:conclu}

In this work, we have formulated FQAs as a new approach for ground state preparation for the Fermi-Hubbard model and more general molecular systems. FQAs operate by assigning values to quantum circuit parameters via a feedback law inspired by QLC, and are thus optimization-free. We have presented simulation results exploring their performance on instances of the Fermi-Hubbard model and on small molecular Hamiltonians, and have observed good convergence to the ground state. We have also discussed similarities and differences between FQAs, adiabatic ground state preparation, and VQE, noting that FQAs are likely to be most useful in settings amenable to the implementation of relatively deep circuits and the estimation of $A_k$ at each step via qubit measurements. We have explored this latter point, and note that the sampling complexity analysis in Sec.~\ref{JBL:sec:apply} indicates that the cost of estimating $A_k$ through repeated measurements scales favorably, while the numerical evidence in Fig.~\ref{JBL:fig:robust} suggests that even very coarse and noisy estimates of $A_k$ suffice to obtain good performance.

This work motivates a variety of future avenues for research. For example, experimental implementations of the FQAs formulated here would enable the impact of hardware noise on algorithm performance to be analyzed, and would allow for benchmarking the capabilities of current quantum devices for executing FQAs, and comparing the results against those from other candidate approaches for ground state preparation. Another extension would be to investigate potential benefits of using FQA parameter values as a starting point to seed further classical optimization in VQE, and to compare this against other techniques for seeding the optimization parameter values. This seeding approach was explored for QAOA using parameters found via FALQON in Ref.~\cite{JBL:falqon}, and may be especially relevant in settings where circuit depth is a limiting factor. Future work developing theoretical aspects of FQAs would also be valuable, e.g., to verify their convergence properties. More broadly, it would be interesting to develop FQAs for other Hamiltonian models, e.g., extensions of the Fermi-Hubbard model \cite{Consiglio_2022} and first-quantized representations of molecular systems, to explore FQAs for minimizing nonlinear cost functions, such as those found in quantum machine learning, to explore FQA formulations that leverage intermediate-time measurements \cite{315a253213bb43839bccbb02502da78a}, and to further develop extensions to imaginary time formulations \cite{PhysRevResearch.5.023087}. Finally, we remark that the adaptation of FQAs to fault-tolerant quantum computing architectures may also be a promising path forward, as the measurement overheads of VQE can come to dominate the cost for certain types of simulation problems \cite{gonthier2022measurements}, even for relatively small system sizes. In closing, we note that this work explores one particular connection between quantum algorithms and quantum control, i.e., through the development of a quantum Lyapunov control-inspired quantum algorithm for ground state preparation. Looking ahead, there may be a variety of other opportunities for ideas from quantum dynamics and control to serve as inspiration for new directions in quantum algorithms research \cite{PhysRevX.7.021027,PRXQuantum.2.010101, PhysRevLett.126.070505, meitei2021gate, PhysRevResearch.3.023092, meirom2022pansatz, Larocca2022diagnosingbarren}.

\begin{acknowledgments}

We are grateful to Lucas Kovalsky, Mohan Sarovar, and Stefan Seritan for insightful conversations and manuscript feedback, and thank Lucas Brady for observing and discussing \bibnotemark[bibnotecorrection]. The authors acknowledge support from Sandia National Laboratories’ Laboratory Directed Research and Development Program under the Truman Fellowship and Project 222396. MDG also acknowledges support from the U.S. Department of Energy, Office of Science, Office of Advanced Scientific Computing Research, under the Accelerated Research in Quantum Computing (ARQC) program. Sandia National Laboratories is a multimission laboratory managed and operated by National Technology \& Engineering Solutions of Sandia, LLC, a wholly owned subsidiary of Honeywell International Inc., for the U.S. Department of Energy’s National Nuclear Security Administration under contract DE-NA0003525. This paper describes objective technical results and analysis. Any subjective views or opinions that might be expressed in the paper do not necessarily represent the views of the U.S. Department of Energy or the United States Government. SAND2023-08906O.

\end{acknowledgments}

\bibliographystyle{quantum}
\bibliography{FQAGSP_arxiv_Sep26.bib}

\begin{thebibliography}{10}
\bibitem{bibnotecorrection}
The reference field perturbation heuristic is motivated in Sec. IIIA of \cite
  {JBL:longfalqon}, which discusses the continuous-time formulation. The first
  line of Eq. (14) in this section should read $\frac {d}{dt}E_{p,(b)} =
  \langle \psi (t)|i [H(t), H_{p,(b)}(t)] |\psi (t)\rangle + \frac {d\lambda
  (t)}{ dt} \langle \psi (t)|H_d |\psi (t)\rangle $, with the following lines
  of the equation updated accordingly. Consequently, the subsequent reasoning
  in Sec. IIIA of \cite {JBL:longfalqon} holds only if $\Big |\frac {d\lambda
  (t)}{ dt} \langle \psi (t)|H_d |\psi (t)\rangle \Big |$ is negligible, e.g.,
  if the variation in the reference field perturbation, $\lambda (t)$, with
  time is sufficiently slow.
\end{thebibliography}

\onecolumn
\appendix

\section{Sampling Complexity Analysis}
\subsection{Fermi-Hubbard Model}
\label{app:fhmod}
In this Appendix, we investigate the cost of estimating $A_k$ of Eq.~\eqref{eq:ak} for the Fermi-Hubbard model through repeated measurements of the observable corresponding to $i[T,V]$. We begin by considering the commutator   
\begin{equation}
    [T,V] = -\tau U\sum_{ij}[a_i^\dagger a_j + a_j^\dagger a_i, \sum_{k\ell} n_kn_\ell],
    \label{eq:comm1}
\end{equation}
where $i,j$ index nearest-neighbor spin orbitals of same spin and $k,\ell$ index spin orbitals of opposite spin associated with the same site. We observe that each term in the commutator from $T$, i.e., $a_i^\dagger a_j + a_j^\dagger a_i$, will have a nonvanishing commutator with two terms from $V$, $n_in_a$ and $n_jn_b$, that are associated with the lattice sites hosting the $i$ and $j$ spin orbitals, respectively, as well as the $a$ and $b$ spin orbitals of the opposite spin. In this notation, Eq.~\eqref{eq:comm1} evaluates to
\begin{equation}
    [T, V] = -\tau U \sum_{i,j} (a_i^\dagger a_j - a_j^\dagger a_i) (n_b - n_a).
\end{equation}
In the Jordan-Wigner encoding,
\begin{equation}
\begin{aligned}
    a_i^\dagger a_j - a_j^\dagger a_i &= \frac{i}{2}(X_iY_j - X_jY_i)Z_{i+1}...Z_{j-1}, \\
    n_b - n_a &= \frac{1}{2}(Z_a-Z_b)
\end{aligned}
\end{equation}
such that 
\begin{equation}
    [T,V] = -\frac{i\tau U}{4}\sum_{i,j} (X_iY_jZ_a - X_jY_iZ_a - X_iY_jZ_b + X_jY_iZ_b)Z_{i+1}...Z_{j-1}.
\end{equation}

\begin{figure}[hb]
    \centering
    \includegraphics[scale=0.175]{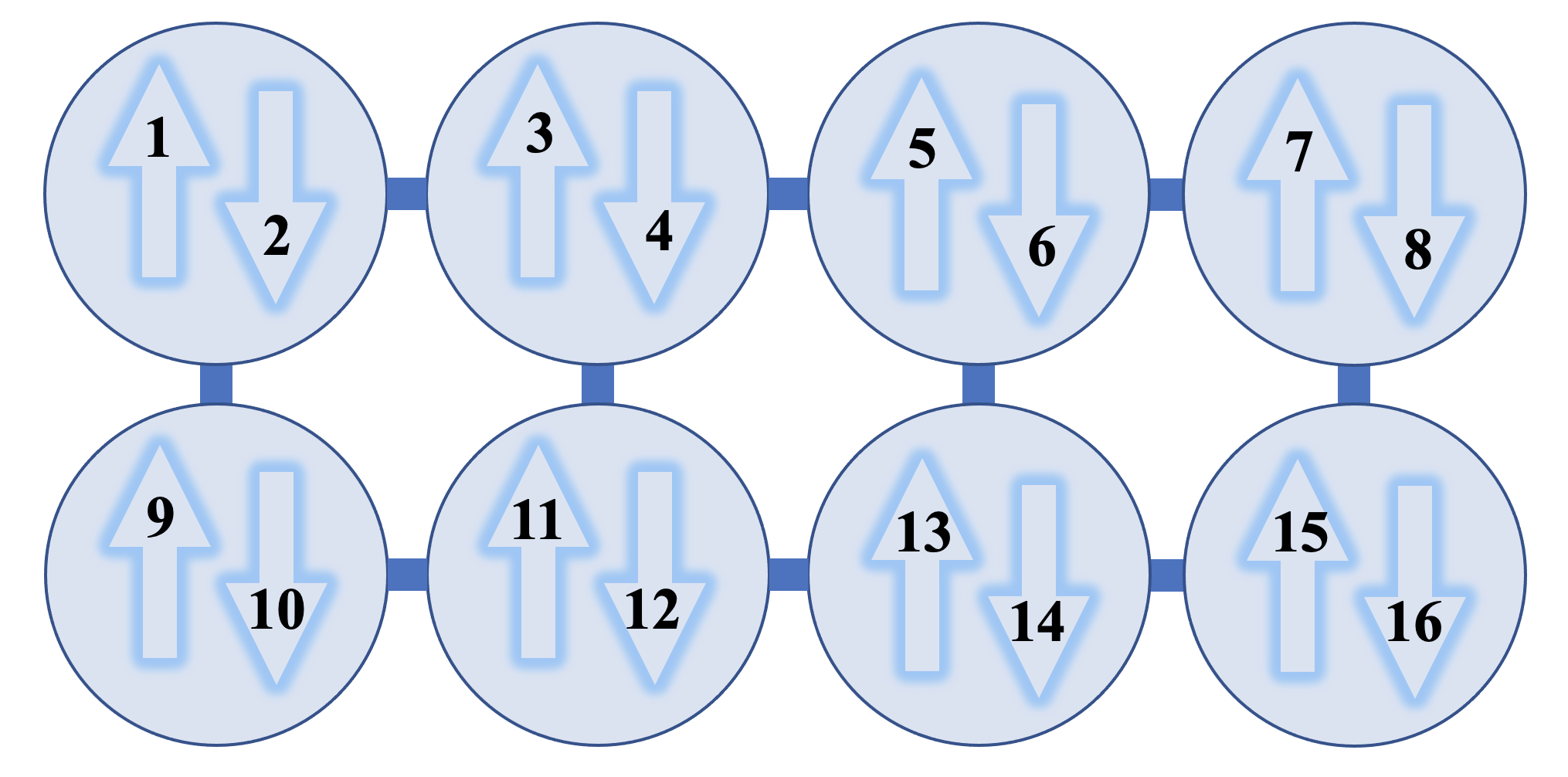}
    \caption{Example fermion-to-qubit labeling scheme, depicted for $2\times 4$ Fermi-Hubbard lattice.}
    \label{qubitnumbering}
\end{figure}

For a generic $n_r \times n_c$ square lattice, where $n_r$ denotes the vertical dimension, i.e., number of rows, and $n_c$ denotes the horizontal dimension, i.e., number of columns, there are $2n_c n_r$ (or $2n_c n_r-n_c-n_r$) nearest-neighbor $i,j$ pairs assuming periodic (or open) boundary conditions. In the following, we restrict our consideration to open boundary conditions, where the number of Pauli strings in Eq.~\eqref{eq:JWencodedcomm} is given by $16n_cn_r-8n_c-8n_r$. In principle, each of these Pauli strings could be considered individually; their expectation values could be estimated using $m$ measurement samples each, and the expectation value $A_k$ could be estimated by taking the appropriate linear combination of the results. This corresponds to the use of $N_{samp} = (16n_cn_r-8n_c-8n_r)m$ measurement samples to estimate $A_k$.

We now consider how measurements of these Pauli strings can be parallelized to reduce the sampling complexity. The specific parallelization depends on the fermion-to-qubit labeling scheme that is used. We discuss one particular formulation using the mapping depicted in Fig.~\ref{qubitnumbering} (i.e., shown for concreteness for a $2\times4$ lattice), noting that similar parallelization techniques can be used for other mappings. We begin by grouping the Pauli strings into three categories: (1) those with weight-two Pauli strings (e.g., $X_1Y_3$ and $Y_6X_8$), (2) those with weight-four Pauli strings, and (3) those with weight-$2n_c$ and weight-($2n_c+2$) Pauli strings. We next observe that we can partition the elements in (1) into two sets, such that the elements within each set commute with one another and can be measured in parallel (e.g., $X_1Y_3, X_2Y_4, Y_3X_5, Y_4X_6, X_5Y_7, X_6Y_8$, etc, comprise the first set and $Y_1X_3, Y_2X_4, X_3Y_5, X_4Y_6, Y_5X_7, Y_6X_8$, etc, comprise the second set). As such, the expectation values of the Pauli strings in (1) can be estimated using $m$ repetitions of 2 measurement circuits, i.e., one measurement circuit per set of mutually commuting terms. Following this same line of reasoning, we observe that the expectation values of the Pauli strings in (2) can be estimated with $m$ repetitions of 4 measurement circuits (e.g., to estimate $X_1Z_2Y_3Z_4, Y_3Z_4X_5Z_6, X_5Z_6Y_7Z_8$, etc; $Y_1Z_2X_3Z_4, X_3Z_4Y_5Z_6$, etc; $Z_1X_2Z_3Y_4, Z_3Y_4Z_5X_6$, etc; and $Z_1Y_2Z_3X_4, Z_3X_4Z_5Y_6$, etc). Thus, the sampling complexity for (1) and (2) does not scale with the lattice dimension $n_r$ or $n_c$. Finally, the expectation values of the $8n_cn_r - 8n_c$ Pauli strings in (3) can be estimated using $4mn_c$ measurements through parallelization, following analogous layering techniques as used to parallelize (1) and (2). As a result, the total number of samples to estimate $A_k$ for an $n_r\times n_c$ lattice is independent of $n_r$ and is given by $N_{samp} = (4n_c+6)m$, representing a significant reduction in sampling complexity through parallelization.

\subsection{Molecular Hamiltonians}
\label{app:molham}
In this Appendix, we derive the cost of estimating $A_k$ in Eq.~\eqref{eq:mhfeedbacklaw} for generic molecular Hamiltonians through repeated measurements of the observable corresponding to $i[H_1,H_2]$. In terms of fermionic creation and annihilation operators, this commutator is given by Eq. (\ref{mhfermopcomm}) in the main text, reproduced here for convenience:
\begin{equation}
\label{app:eq:h1h2}
[H_1, H_2] = \frac{1}{2} \sum_{ij} t_{ij} \sum_{abcd} u_{abcd} [(a_i^\dagger a_b^\dagger \delta_{ja} + a_a^\dagger a_i^\dagger \delta_{jb})a_c a_d + a_a^\dagger a_b^\dagger (a_j a_d \delta_{ic} + a_c a_j \delta_{id})].
\end{equation}
In the Jordan-Wigner encoding, the annihilation and creation operator associated with the $n$-th spin orbital, i.e., $a_n$ and $a_n^\dagger$, are mapped to $X+iY$ and $X-iY$ acting on qubit $n$, respectively, and Pauli-$Z$ operators acting on remaining qubits. 

As in Sec.~\ref{app:fhmod}, we now proceed to group the various Pauli strings in the commutator into sets of mutually commuting terms that can be measured in parallel, and denote the number of measurement samples used to estimate a given expectation value by $m$. Each fermionic operator on the right-hand side of Eq.~\eqref{app:eq:h1h2} takes the form of $a_a^\dagger a_b^\dagger a_c a_d$. Since the sum in Eq.~\eqref{app:eq:h1h2} iterates over every combination of these four indices up to $n$, there are three cases to consider: 
\begin{enumerate}
    \item Both $a=c$ and $b=d$, so these fermionic operators can be written $a_a^\dagger a_b^\dagger a_a a_b$. The Jordan-Wigner encoding of these operators consists of identity or Pauli-$Z$ operators acting on every qubit, and therefore form a set of mutually commuting terms. As such, the expectation values of these terms can be estimated together with $m$ measurements.
    \item Either $a=c$ or $b=d$, but not both. Without loss of generality, we can let $a=c$ and $b \neq d$, so we are considering fermionic operators of the form $a_a^\dagger a_b^\dagger a_a a_d$. The Jordan-Wigner encoding will result in Pauli strings containing $X+iY$ ($X-iY$) on qubit $d$ ($b$), and $Z$ or $I$ on remaining qubits. Therefore, with each choice of two indices $b$ and $d$ among the $n$ qubits, there are four different sets of mutually commuting terms corresponding to the Pauli strings containing $X_bX_d$, $X_bY_d$, $Y_bX_d$, and $Y_bY_d$. Hence, we can estimate the expectation values of these Pauli strings in $4 m \binom{n}{2}$ measurements.
    \item $a \neq c$ and $b \neq d$. In this case, none of the four creation/annihilation operators cancel each other out, resulting in $2^4 \binom{n}{4}$ different sets. Each choice of four qubits $a,b,c,d$ has $2^4=16$ mutually commuting sets of terms, corresponding to every combination of $X$ and $Y$ acting on these four qubits. We note that this case is only relevant for $n \geq 4$ qubits. 
\end{enumerate}
Combining the sets of mutually commuting terms from the above three cases, we get that the total number of samples required to estimate $A_k$ for an arbitrary molecular Hamiltonian is
\begin{equation}
m\left(1 + 2^2 \binom{n}{2} + 2^4 \binom{n}{4}\right) = m(\frac{2}{3}n^4 - 4n^3 + \frac{28}{3}n^2-6n+1).
\end{equation}

\section{Extension to Multiple Parameters per Layer}
\label{JBL:appendixmcf}
\begin{figure}[htb]
\begin{center}
\scalebox{0.29}{\includegraphics{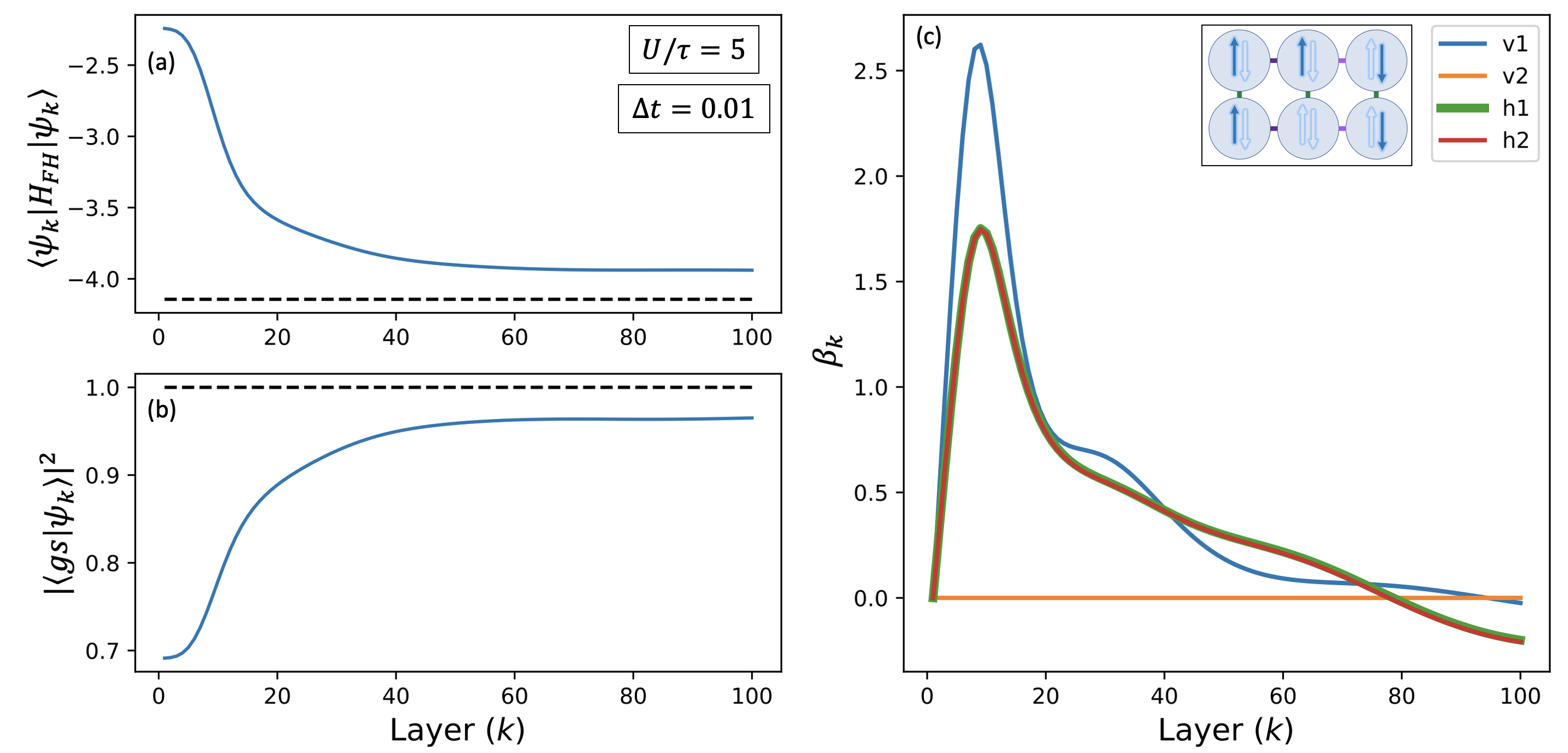}} \caption{Numerical simulation results for the multiparameter FQA applied to a $2 \times 3$ Fermi-Hubbard lattice with $3\uparrow 2\downarrow$ fermions for $\Delta t=0.01$ and $U/\tau=5$.}\label{JBL:fig:multparam}
\end{center}\end{figure}

The decomposition in Eq.~\eqref{JBL:eq:oneparamhva} suggests a natural alternative formulation for an FQA, where each layer contains multiple parameters, as in \cite{JBL:JWFermiEnc}. Here, we develop a multiparameter formulation using the generalization of QLC to the case of multiple control functions, as outlined in Sec.~III of \cite{JBL:longfalqon} and references therein. In particular, beginning with a continuous-time picture, we separate $H_d = T$ into four separate terms $H_{h1},H_{h2},H_{v1},H_{v2}$, to be controlled separately via associated control functions, such that the time evolution is given by
\begin{equation}
    i\frac{d}{dt}|\psi(t)\rangle = \big(H_p +\sum_{j } \beta^j(t)H_j\big)|\psi(t)\rangle,
\end{equation}
for $j \in \{ v2,h2,v1,h1\}$, where $\beta^j(t)$ denotes the time-dependent control function associated with $H_j$, leading to the following modification of Eq.~\eqref{JBL:eq:changeinexp}:
\begin{align}
\frac{dJ}{dt} = \sum_{j \in \{ v2,h2,v1,h1\}} \langle \psi(t)| i[H_j,H_{FH}]|\psi(t) \rangle \beta^j(t).
\label{JBL:eq:multiparamchangeexp}
\end{align}
To ensure Eq.~\eqref{JBL:eq:multiparamchangeexp} is nonpositive, we can set our feedback law in the digitized picture according to a generalization of Eq.~\eqref{JBL:eq:feedbacklaw}, yielding
\begin{equation}
\beta_{k}^j = -\langle \psi_{k-1}| i[H_j,H]|\psi_{k-1} \rangle.
\label{JBL:eq:multiparamfeedbacklaw}
\end{equation}
This yields the following modified formulation for each layer:
\begin{equation}
|\psi_{k+1}\rangle = U_d(\beta^{v2}_k,\beta^{h2}_k,\beta^{v1}_k,\beta^{h1}_k)U_p|\psi_k\rangle,
\end{equation}
where $U_p$ retains the same form as in Eq.~\eqref{JBL:eq:problemunitary}, and 
\begin{equation}
U_d(\beta^{v2}_k,\beta^{h2}_k,\beta^{v1}_k,\beta^{h1}_k) = e^{-i\beta^{h1}_kH_{h1}\Delta t}e^{-i\beta^{v1}_kH_{v1}\Delta t}e^{-i\beta^{h2}_kH_{h2}\Delta t}e^{-i\beta^{v2}_kH_{v2}\Delta t}.
\end{equation}
The associated quantum circuit diagrams are given in Fig.~\ref{fig:fh}(c). This formulation takes advantage of the additional decomposition of $U_d$, but does not necessarily guarantee better performance than the single parameter formulation. In Fig.~\ref{JBL:fig:multparam}, we present a numerical simulation using this multiparameter formulation for a 2$\times$3 Fermi-Hubbard lattice, noting that since the vertical dimension we consider is 2, the $\beta_k^{v2}$-parameters are zero at every layer. 
We observe that despite the additional parameters introduced, the performance in Fig.~\ref{JBL:fig:multparam} is slightly worse than the performance in Fig.~\ref{fig:fhres}(d-f), which shows results for the same Fermi-Hubbard lattice using the single parameter formulation. This may be due to the fact that Eq.~\eqref{JBL:eq:multiparamchangeexp} is overconstrained, i.e., in order for $\frac{dJ}{dt}$ to be nonpositive, there is no requirement that each of the terms in the sum in Eq.~\eqref{JBL:eq:multiparamchangeexp} are negative, only that the full summation is. This indicates there may be benefits to further investigation into multiparameter formulations and the development of other associated feedback laws.

\end{document}